\documentclass[aip,reprint,amsmath,amssymb]{revtex4-1}

\usepackage{graphicx}
\usepackage{dcolumn}
\usepackage{float}
\usepackage{bm}
\usepackage{amssymb}
\usepackage{lipsum}

\begin{document}

\newcommand{\vrr}{\mbox{$\bf {r}$}}
\newcommand{\vRR}{\mbox{$\bf {R}$}}
\newcommand{\be}{\begin{eqnarray}}
\newcommand{\ee}{\end{eqnarray}}
\newcommand{\bes}{\begin{eqnarray*}}
\newcommand{\ees}{\end{eqnarray*}}
\newcommand{\nn}{\nonumber}

\title{Free-energy cost of localizing a single monomer of a confined polymer}

\author{James M. Polson and Zakary R. N. McLure}
\affiliation{ Department of Physics, University of Prince Edward Island, 
550 University Ave., Charlottetown, Prince Edward Island, C1A 4P3, Canada }

\date{\today}

\begin{abstract}
We describe a simple Monte Carlo simulation method to calculate the free-energy cost of localizing 
a single monomer of a polymer confined to a cavity. The localization position is chosen to be
on the inside surface of the confining cavity. The method is applied to a freely-jointed hard-sphere 
polymer chain confined to cavities of spherical and cubic geometries. In the latter case we consider 
localization at a corner and at the center of a face of the confining cube.
We consider cases of end-monomer localization both with and without tethering of the other
end monomer to a point on the surface. We also examine localization of monomers at arbitrary
position along the contour of the polymer. We characterize the dependence of the free energy 
on the cavity size and shape, the localization position, and the polymer length. The quantitative 
trends can be understood using standard scaling arguments and use of a simple theoretical model. 
The results are relevant to those theories of polymer translocation that focus on the importance
of the free-energy barrier as the translocation process requires an initial localization of a 
monomer to the position of a nanopore.
\end{abstract}

\maketitle

\section{Introduction}
\label{sec:intro}

The translocation of polymers through nanopores into, out of, or between enclosed spaces has
been the subject of much theoretical interest for many years.%
\cite{muthukumar1989effects,muthukumar2001translocation,muthukumar2003polymer,kong2004polymer,%
polson2015polymer,%
magill2018sequential,laachi2010statistics,rasmussen2012translocation,saltzman2009conformation,%
dell2018anomalous}
This work is driven in part by its relevance to various biological phenomena such as the
packing and ejection of viral DNA as well as the transport of mRNA through the nuclear pore 
complex.\cite{Alberts_book,Lodish_book}
Another key motivation is for modeling the structural and dynamical behaviour of polymers in 
randomly structured porous environments such as those composed of polymer networks and gels. 
In the so-called barrier-dominated regime, the chain dynamics is controlled by entropic trapping 
in and escape from cavity-like spaces separated by narrow constrictions. 
This behavior has also been studied in ordered media patterned with molecular-sized spatial
constraints fabricated using colloidal templating.\cite{nykypanchuk2002brownian}
Modeling such systems as a collection of spherical chambers separated by narrow cylindrical pores,%
\cite{saltzman2009conformation,dell2018anomalous,laachi2010dna} computer simulations
have revealed a rich variety of dynamical behavior that is governed largely by the
relative size of the polymer and the chambers and the resulting spatial distribution
of the polymer. If the chamber/polymer size ratio is small, the polymer may typically span 
several cavities, while for larger ratios the entire polymer resides in a single chamber and 
occasionally translocates wholly into a neighboring one. Another related simulation study%
\cite{magill2016translocation} examined similar effects for translocation through a simple 
two-nanopore single-cavity structure whose design was motived by previous experimental work.%
\cite{langecker2011electrophoretic} More recently, Magill {\it et al.} extended this work
and examined the translocation behavior of polymers in a sequential nanopore-channel device,
which was proposed as a means to separate polymers by size.\cite{magill2018sequential} 

Central to some theories of dynamics of polymers in porous media is the concept of 
an entropic barrier.\cite{muthukumar1989effects,Muthukumar_book} The passage of the
polymer through a pore imposes spatial constraints that significantly reduce its conformational
entropy and thus increase the free energy. Since the height of the barrier is the main
determinant of the dynamical behavior in the barrier-dominated regime, its characterization
with respect to the system properties is essential. There are two main contributions 
to the barrier. The first is the free-energy cost of localizing an end monomer to the 
location of a pore, and the second is determined by the variation of the free energy
with respect to the degree of translocation during the so-called threading stage.%
\cite{Muthukumar_book} The second contribution has been extensively studied. 
Analytical approximations for the free-energy function were prominent in some of
the earliest theoretical studies of translocation,\cite{sung1996polymer,muthukumar2001translocation}
and its main scaling properties have have recently been characterized using Monte Carlo simulations.%
\cite{polson2013simulation,polson2015polymer}

The contribution to the free-energy barrier from end-monomer localization has received less 
attention.  Analytical expressions for this free energy have been derived by Muthukumar and coworkers 
using the Green's function solution for a Gaussian chain under spherical confinement with
and without tethering of an end monomer to the confining surface.\cite{muthukumar2003polymer,%
kong2004polymer,dell2018anomalous} However, there have been no corresponding calculations
using simulation methods carried out to test the accuracy of such analytical results. 
Two previously developed simulation techniques can in principle be used for this purpose.
The first is that of Laachi and Dorfman, who introduced a Monte Carlo method to estimate the 
partition function (and therefore total free energy) of single- and double-tethered lattice-model 
polymers confined to a spherical cavity.\cite{laachi2010statistics}  
Another approach is the incremental gauge cell method developed by Rasmussen {\it et al.},\cite{%
rasmussen2011calculation} which was used to calculate the incremental chemical potential (and 
thus the total confinement free energy) of a chain tethered to the inner surface of a confining 
sphere.\cite{rasmussen2012translocation} The end-monomer localization free energy can be 
calculated as the difference between the free energies of tethered and untethered polymers
confined to a cavity, where the latter quantity can be calculated using a technique such as 
thermodynamic integration.\cite{cacciuto2006self} 

In this study we describe a Monte Carlo simulation method to calculate the free-energy cost of 
localizing a single monomer to the inner surface of a confinement cavity. The method has 
several advantages over the two referenced above. First, it does not require a separate
calculation of the confinement free energy of an untethered chain. In addition, the simulations
do not require use of a lattice model as in Ref.~\onlinecite{laachi2010statistics}. Like
the approach in Ref.~\onlinecite{laachi2010statistics} and unlike that of
Ref.~\onlinecite{rasmussen2012translocation} it can easily be used to determine the end-monomer
localization free energy in the case where the other end monomer is tethered to a different
position in the cavity. This feature is essential 
when using the results to understand the polymer dynamics in systems where the polymer spans 
several cavities.\cite{saltzman2009conformation,dell2018anomalous,laachi2010dna}
Finally, our method can be used to calculate the free-energy cost of localizing {\it any}
monomer for a polymer of {\it arbitrary} topology (e.g. linear, ring, branched). Such 
versatility enables calculations relevant for example to the translocation of a folded polymer, 
where the first monomer through the pore is not an end monomer.\cite{mihovilovic2013statistics}

We use the method to calculate the localization free energy for the simple but illustrative cases
of a single linear polymer confined to a spherical and a cubic cavity. In the latter case, we 
consider end-monomer localization to both the center of one cube face and to a corner of the cube. 
We also examine localization for monomers at various locations along the chain contour, as well
as the case of end-monomer localization for a chain with one tethered end. We show that the 
results are quantitatively consistent with known results in the limiting case of very weak
confinement. Using a simple theoretical model we also show that the scaling of the free energy
with polymer length and cavity size and shape is semi-quantitatively consistent with the expected
trends.

The remainder of this article is organized as follows. In Section~\ref{sec:theory} we present
and justify the Monte Carlo (MC) algorithm for calculating the free energy. 
Section~\ref{sec:model} provides
a brief description of the model employed in the simulations, while Section~\ref{sec:methods} 
provides the relevant details of the implementation of the algorithm in the simulation.
Section~\ref{sec:results} presents the simulation results for the various systems we have
examined, as well as the predictions from theoretical models developed in the appendices.
Finally, Section~\ref{sec:conclusions} summarizes the main conclusions of this work.

\section{Development of the method}
\label{sec:theory}

In this section we develop the Monte Carlo method used to calculate the free-energy cost of 
localizing a single monomer to a point on the inside surface of a cavity within which the 
polymer is confined. For convenience, we present the theoretical justification of the algorithm
for the special case of localizing an end monomer in the absence of any constraints other 
than confinement to the cavity. 

\subsection{Free-energy cost of partial localization of chain end}
\label{subsec:partial}

We first describe a method to measure the free-energy difference between two polymer systems 
with difference degrees of confinement of a single end monomer. In one system the end monomer
is confined to a volume $V^\prime_{\rm a}$, while the other is confined to a volume 
$V^\prime_{\rm b}$, which is chosen to be a subvolume of $V^\prime_{\rm a}$. In addition, 
both $V^\prime_{\rm a}$ and $V^\prime_{\rm b}$ are subvolumes of the volume $V$ of the cavity 
within which the entire polymer is confined.

Consider a freely-jointed polymer chain of $N+1$ monomers with positions 
$\vrr_0,\vrr_1,\vrr_2,...,\vrr_N$. We define the related set of $N+1$ vector coordinates
$\vRR_n$ such that $\vRR_0 = \vrr_0$ and $\vRR_n = \vrr_n-\vrr_{n-1}$ for $n$ ranging from 1 
to $N$. Consider as well the case where the polymer is confined to a box of volume $V$ 
and one end monomer is confined to a volume $V^\prime<V$, where $V^\prime$ lies within 
the volume $V$.  The potential energy of the polymer, 
$U(\vRR_0,\vRR_1,...,\vRR_N;V^\prime) \equiv U(\vRR_0,\vRR^N;V^\prime)$, can be partitioned,
\be
U = U_{\rm int}(\vRR^N) +  U_{\rm ext}(\vRR_0,\vRR^N;V'),
\ee
where $U_{\rm int}(\vRR^N)$ is the internal potential energy associated with the 
interactions between bonded and non-bonded monomers, and $U_{\rm ext}(\vRR_0,\vRR^N;V')$
is the interaction between the monomers and the confining walls.
The configurational partition function is given by
\be
 Z(V') & = & \int d^3\vRR_0\int d^{3N}\vRR \exp\left[-\beta (U_{\rm int}(\vRR^N) \right.\nonumber\\
      &   & \left. + U_{\rm ext}(\vRR_0,\vRR^N;V'))\right],
\label{eq:ZVp}
\ee
and the configurational contribution to the free energy is
\be
F(V') = -k_{\rm B}T\ln Z(V'),
\label{eq:FVp}
\ee
where $k_{\rm B}$ is Boltzmann's constant and $T$ is absolute temperature.

Now consider two systems, $\bf{a}$ and $\bf{b}$, distinguished only by different 
containment volumes for the $n$=0 monomer, $V'_{\rm a}$ and $V'_{\rm b}$.
The difference in the free energies for these states is
\be
\Delta F \equiv F(V'_{\rm b})-F(V'_{\rm a}) = -k_{\rm B}T 
\ln\left[\frac{Z(V'_{\rm b})}{Z(V'_{\rm a})}\right].
\ee
Thus,
\begin{widetext}
\bes
\beta \Delta F & = & -\ln\left[ \frac {\int d^{3(N+1)}\vRR \exp\left[-\beta U_{\rm int}(\vRR^N)\right]
\exp\left[-\beta U_{\rm ext}(\vRR^{N+1};V'_{\rm b})\right] } 
{\int d^{3(N+1)}\vRR \exp\left[-\beta U_{\rm int}(\vRR^N)\right] 
\exp\left[-\beta U_{\rm ext}(\vRR^{N+1};V'_{\rm a})\right] } \right],
\ees
\end{widetext}
where $\beta\equiv 1/k_{\rm B}T$.
Let the confinement volume for case {\bf b} lie within the confinement volume for case {\bf a}.
Obviously, this also implies that $V_{\rm b}<V_{\rm a}$. In addition, it is clear that
\bes
 \exp\left[-\beta U_{\rm ext}(\vRR^{N+1};V'_{\rm b})\right] =
 \exp\left[-\beta (U_{\rm ext}(\vRR^{N+1};V'_{\rm b}) \right.  \\
  \left. + U_{\rm ext}(\vRR^{N+1};V'_{\rm a}))\right].
\ees
This follows from the fact that if the end monomer lies with $V'_{\rm b}$, where 
$U_{\rm ext}(\vRR^{N+1};V'_{\rm b})=0$, then it also lies within the volume $V'_{\rm a}$,
where $U_{\rm ext}(\vRR^{N+1};V'_{\rm a})=0$. Likewise, if it lies outside $V'_{\rm b}$,
then $U_{\rm ext}(\vRR^{N+1};V'_{\rm b})=\infty$, regardless of whether it lies inside
or outside of $V'_{\rm a}$. Thus, 
\begin{widetext}
\be
\beta\Delta F & = & 
          -\ln\left[ \frac {\int d^{3(N+1)}\vRR
           \exp\left[-\beta U_{\rm ext}(\vRR^{N+1};V'_{\rm b})\right] 
           \exp\left[-\beta U(\vRR^{N+1};V'_{\rm a})\right] }
           {\int d^{3(N+1)}\vRR \exp\left[-\beta U(\vRR^{N+1};V'_{\rm a})\right] } \right]
\ee
\end{widetext}
or
\be
\beta\Delta F & = & -\ln\left\langle \exp\left[-\beta U_{\rm ext}(\vRR^{N+1};V'_{\rm b})\right] 
\right\rangle_{a},
\label{eq:bDF}
\ee
where $\langle\cdots\rangle_{\rm a}$ denotes an ensemble average using the potential 
$U(\vRR^{N+1},V'_{\rm a})$.

To estimate this free-energy difference in a MC simulation, the configurational states 
of the system are generated using the potential $U(\vRR^{N+1},V'_{\rm a})$.  Thus, the 
end monomer is confined to the volume $V^\prime_{\rm a}$ while the remaining monomers are confined 
to the volume $V$. The average is obtained by sampling the exponential in Eq.~(\ref{eq:bDF}) 
using these configurations.  The external potential energies of monomers labeled by $n$=1 
to $N$ will always be zero, since these monomers always lie within the volume $V$.  On the 
other hand, the external potential energy for the $n$=0 monomer is $u_{\rm ext}(\vRR_0)=0$ 
if ${\bf R}_0$ is inside $V'_{\rm b}$ and $u_{\rm ext}(\vRR_0)= \infty$, if ${\bf R}_0$ lies 
outside $V'_{\rm b}$. Thus,
\be
\exp\left[-\beta U_{\rm ext}(\vRR^{N+1};V'_{\rm b})\right] & = & 
1, ~~ {\rm if}~{\bf R}_0~ {\rm inside~}  V'_{\rm b} \nonumber\\
& = & 0, ~~ {\rm if}~{\bf R}_0~ {\rm outside~} V'_{\rm b}
\label{eq:expUext}
\ee
If this quantity is sampled in a MC simulation $M$ times and the $m$th sampling gives a 
value $S_m$ then the free-energy difference is then approximately 
\begin{eqnarray}
\beta\Delta F \approx -\ln\left[ (1/M) \sum_{m=1}^M S_m \right]. 
\label{eq:Sm}
\end{eqnarray}
The approximation is expected to become exact in the limit $M\rightarrow\infty$.

\subsection{Localization of the end monomer}
\label{subsec:localization}

We now show how the procedure described above in Section~\ref{subsec:partial} to measure
the free-energy difference between systems of different degrees of confinement of
the chain end can be used to measure the free-energy cost of localizing the end monomer
to a point on the confinement surface.
Returning to the configurational partition function of Eq.~(\ref{eq:ZVp}), we write
\bes
 Z(V') & = & \int d^{3N}\vRR \exp\left[-\beta U_{\rm int}(\vRR^N)\right] \nonumber \\
 && \times \int d^3\vRR_0 \exp\left[-\beta U_{\rm ext}(\vRR_0,\vRR^N;V')\right].
\ees
Defining the quantity
\bes
W_{\rm ext}(\vRR^N,V') = -k_{\rm B}T\ln \Omega(\vRR^N,V'),
\ees
where 
\bes
V' \Omega(\vRR^N,V') \equiv \int d^3\vRR_0 \exp\left[-\beta U_{\rm ext}(\vRR_0,\vRR^N;V')\right],
\ees
it follows
\bes
Z(V') & = & V' \int d^{3N}\vRR \exp\left[-\beta (U_{\rm int}(\vRR^N)+W_{\rm ext}(\vRR^N,V'))\right].
\ees
Next, we define the contribution from the internal energy to the configurational partition
function:
\bes
Z_{\rm int}(V') \equiv \int d^{3N}\vRR 
\exp\left[-\beta (U_{\rm int}(\vRR^N)+W_{\rm ext}(\vRR^N,V'))\right],
\ees
from which it follows 
\be
Z(V') = V' Z_{\rm int}.
\label{eq:ZVp2}
\ee
The configurational free energy is given by
\bes
F(V') =  -k_{\rm B}T\ln V' + F_{\rm int}(V'),
\ees
where we have defined
\bes
F_{\rm int}(V') \equiv -k_{\rm B}T \ln Z_{\rm int}(V').
\ees
The quantity $F_{\rm int}(V')$ can be considered the conformational free energy
for a polymer with an end monomer constrained to lie within $V'$.
Clearly, the free-energy difference between systems with $V'=V'_{\rm a}$ and
$V'=V'_{\rm b}<V'_{\rm a}$ is
\be
\Delta F & \equiv & \left[F(V'_{\rm b})-F(V'_{\rm a})\right] \nonumber \\
& = & k_{\rm B}T \ln\left[V'_{\rm a}/V'_{\rm b}\right] 
+ F_{\rm int}(V_{\rm b}')-F_{\rm int}(V_{\rm a}').
\label{eq:delFab}
\ee

Now consider the case where $V' \rightarrow \delta V$, i.e. the end-monomer confinement volume
becomes very small.  Choose its location to be any point inside $V$ and call this point the origin, 
${\bf 0}$.  For a sufficiently small volume,
\bes
\delta V \Omega(\vRR^N,\delta V,{\bf 0}) & = & \int d^3\vRR_0 \exp\left[-\beta U_{\rm ext}(\vRR_0,\vRR^N;V')\right] \nonumber\\
& \approx &  \delta V  \exp\left[-\beta U_{\rm ext}({\bf 0},\vRR^N)\right]
\ees
and thus
\bes
\Omega(\vRR^N,\delta V,{\bf 0}) & \approx & \exp\left[-\beta U_{\rm ext}({\bf 0},\vRR^N)\right],
\ees
where $U_{\rm ext}({\bf 0},\vRR^N)$ denotes the potential energy of confinement for the polymer 
with the $n$=0 monomer tethered to the origin. The approximation becomes exact in the
limit where $\delta V\rightarrow 0$. It follows that
\bes
W_{\rm ext} (\vRR^N,\delta V) \approx U_{\rm ext}({\bf 0},\vRR^N),
\ees
and thus Eq.~(\ref{eq:ZVp2}) becomes
\bes
Z(\delta V) \approx \delta V\int d^{3N}\vRR 
\exp\left[-\beta (U_{\rm int}(\vRR^N)+U_{\rm ext}({\bf 0},\vRR^N))\right].
\ees
This can be written
\bes
Z(\delta V) \approx \delta V Z^{(0)}_{\rm int},
\ees
where
\bes
Z^{(0)}_{\rm int} \equiv \int d^{3N}\vRR \exp\left[-\beta (U_{\rm int}(\vRR^N)+U_{\rm ext}
({\bf 0},\vRR^N))\right]
\ees
is the configurational partition function for a polymer confined and end-tethered to 
a point inside $V$.  Thus, for small $\delta V$, $Z_{\rm int}(\delta V)\approx Z_{\rm int}^{(0)}$.
Consequently, the configurational free energy satisfies
\bes
F(\delta V)\equiv -kT_{\rm B}\ln\delta V 
+ F_{\rm int}(\delta V)\approx -k_{\rm B}T\ln \delta V + F^{(0)}_{\rm int},
\ees
where
\bes
F^{(0)}_{\rm int} \equiv -k_{\rm B}T\ln Z^{(0)}_{\rm int}
\ees
is the configurational free energy for a case of a polymer tethered at the origin.  
Denoting the difference in the configurational free energy for systems with $V'=V$ 
and $V'=\delta V$  as
\bes
\Delta F(\delta V)\equiv F(\delta V)-F(V)
\ees
it follows that
\be
\Delta F(\delta V) = k_{\rm B}T \ln(V/\delta V) + \Delta F_{\rm int}(\delta V),
\label{eq:FdV}
\ee
where
\bes
\Delta F_{\rm int}(\delta V) \equiv F_{\rm int}(\delta V)-F_{\rm int}(V)
\ees
is approximately
\bes
\Delta F_{\rm int}(\delta V) \approx F^{(0)}_{\rm int} - F_{\rm int}(V).
\ees
This approximation becomes exact in the limit $\delta V\rightarrow 0$. 
We define $\Delta F_{\rm loc}$ as follows:
\be
\Delta F_{\rm loc} & \equiv & \lim_{\delta V\rightarrow 0} \Delta F_{\rm int}(\delta V) \nn \\
& = & \lim_{\delta V\rightarrow 0} \left[ \Delta F(\delta V) - k_{\rm B}T \ln(V/\delta V) \right],
\label{eq:Floc}
\ee
where the difference is expected to converge in the limit. We identify $\Delta F_{\rm loc}$ as 
the chain-end localization free energy. This is the central quantity of this study.

The physical meaning of the quantity $\Delta F_{\rm loc}$ is as follows. Let ${\cal P}(V')$
be the probability that the end monomer lies in the sub-volume $V'$, where $0<V'\leq V$. 
If we choose $V'=\delta V$ and $V'=V$ and note that ${\cal P}(V)=1$, it is clear that
\be
{\cal P}(\delta V) = \frac{{\cal P}(\delta V)}{{\cal P}(V)} 
= \frac{\exp[-\beta F(\delta V)]}{\exp[-\beta F(V)]}.
\ee
Using Eqs.~(\ref{eq:FdV}) and (\ref{eq:Floc}), it is easily shown that
\be
{\cal P}(\delta V) = \left(\frac{\delta V}{V}\right) \exp[-\beta\Delta F_{\rm loc}]
\label{eq:PV}
\ee
for an infinitesimal volume $\delta V$. The volume ratio prefactor is the probability that 
a single isolated particle lies inside $\delta V$ in the absence of other monomers. 
The Boltzmann factor containing 
$\Delta F_{\rm loc}$ is the factor by which this probability is altered by virtue of the 
connection of the end monomer to the rest of the polymer. The case $\Delta F_{\rm loc}>0$ 
corresponds to a probability depletion relative to the isolated-particle case, while 
$\Delta F_{\rm loc}<0$ corresponds to an enhancement in the probability.

In principle, we can calculate $\Delta F_{\rm loc}$ by carrying out a simulation using the 
procedure outlined in Section~\ref{subsec:partial} to measure $\Delta F$ for a small but finite 
volume $\delta V$.  However, for large $V/\delta V$, the statistics are expected to be poor. 
To circumvent this problem we carry out a series of simulations employing a sequence 
of ever-decreasing values of $V'$. Equation~(\ref{eq:delFab}) can be used to calculate the 
free-energy difference for systems of different monomer confinement volumes and the differences 
summed to obtain the chain-end localization free energy. As an example, consider the free 
energy difference between systems with $V'_{\rm a}=V$ and $V'_{\rm b}=V/2$. From Eq.~(\ref{eq:delFab}),
it follows that
\bes
F(V/2) - F(V) & = & k_{\rm B}T \ln 2 + F_{\rm int}(V/2)-F_{\rm int}(V)
\ees
or
\bes
F_{\rm int}(V/2)-F_{\rm int}(V) & = & \Delta F_1 - k_{\rm B}T\ln 2
\ees
where we define $\Delta F_1 \equiv F(V/2)-F(V)$. We can likewise find comparable expressions 
choosing $V'_{\rm a}=V/2$ and $V'_{\rm b}=V/4$, and then $V'_{\rm a}=V/4$ and $V'_{\rm b}=V/8$, 
and so on. Using those results, it is easily shown that
\be
F_{\rm int}(V/2^n) - F_{\rm int}(V) = \sum_{m=1}^n \Delta F_m - n k_{\rm B}T \ln 2,
\ee
where
\be
\Delta F_m \equiv F(V/2^m)-F(V/2^{m-1}).
\label{eq:FmV2}
\ee
Now, for sufficiently large $n$, $\delta V_n = V/2^n$ will be small enough to satisfy the 
previous approximations.  Thus,
\be
F_{\rm int}(V/2^n) - F_{\rm int}(V) \approx \Delta F_{\rm loc},
\ee
or
\be
\Delta F_{\rm loc} \approx \sum_{m=1}^n \left(\Delta F_m - k_{\rm B}T \ln 2\right),
\label{eq:DFint}
\ee
where the approximation becomes more accurate as $n$ increases. 

These results can be generalized by considering the case where the sequence of decreasing 
end-monomer confinement volumes is not generated by subsequent divisions by a factor of 2. 
Instead, we choose arbitrary ratios between successive subvolumes in the sequence. In this case, 
it can be shown that Eq.~(\ref{eq:FmV2}) should be modified to
\be
\Delta F_m \equiv F(V_m)-F(V_{m-1}),
\label{eq:FmV2_2}
\ee
where 
\be
V_{m}\equiv \alpha_m V_{m-1} = \left(\prod_{i=1}^m \alpha_i\right) V,
\label{eq:Vmm}
\ee
and where the set of proportionality constants $\{\alpha_m\}_{m=1}^n$ are arbitrary.
In addition, Eq.~(\ref{eq:DFint}) should be modified to be
\be
\Delta F_{\rm loc} \approx \sum_{m=1}^n \left(\Delta F_m - k_{\rm B}T \ln \alpha_m^{-1}\right).
\label{eq:DFint2}
\ee
In practice, the constants $\{\alpha_m\}_{m=1}^n$ should be chosen in a manner to optimize 
the statistical efficiency of the simulations. Note that the choice $\alpha_{m}=\frac{1}{2}$
reduces Eq.~(\ref{eq:FmV2_2}) to Eq.~(\ref{eq:FmV2}) and Eq.~(\ref{eq:DFint2}) to 
Eq.~(\ref{eq:DFint}).

\begin{figure}[!ht]
\begin{center}
\vspace*{0.2in}
\includegraphics[width=0.45\textwidth]{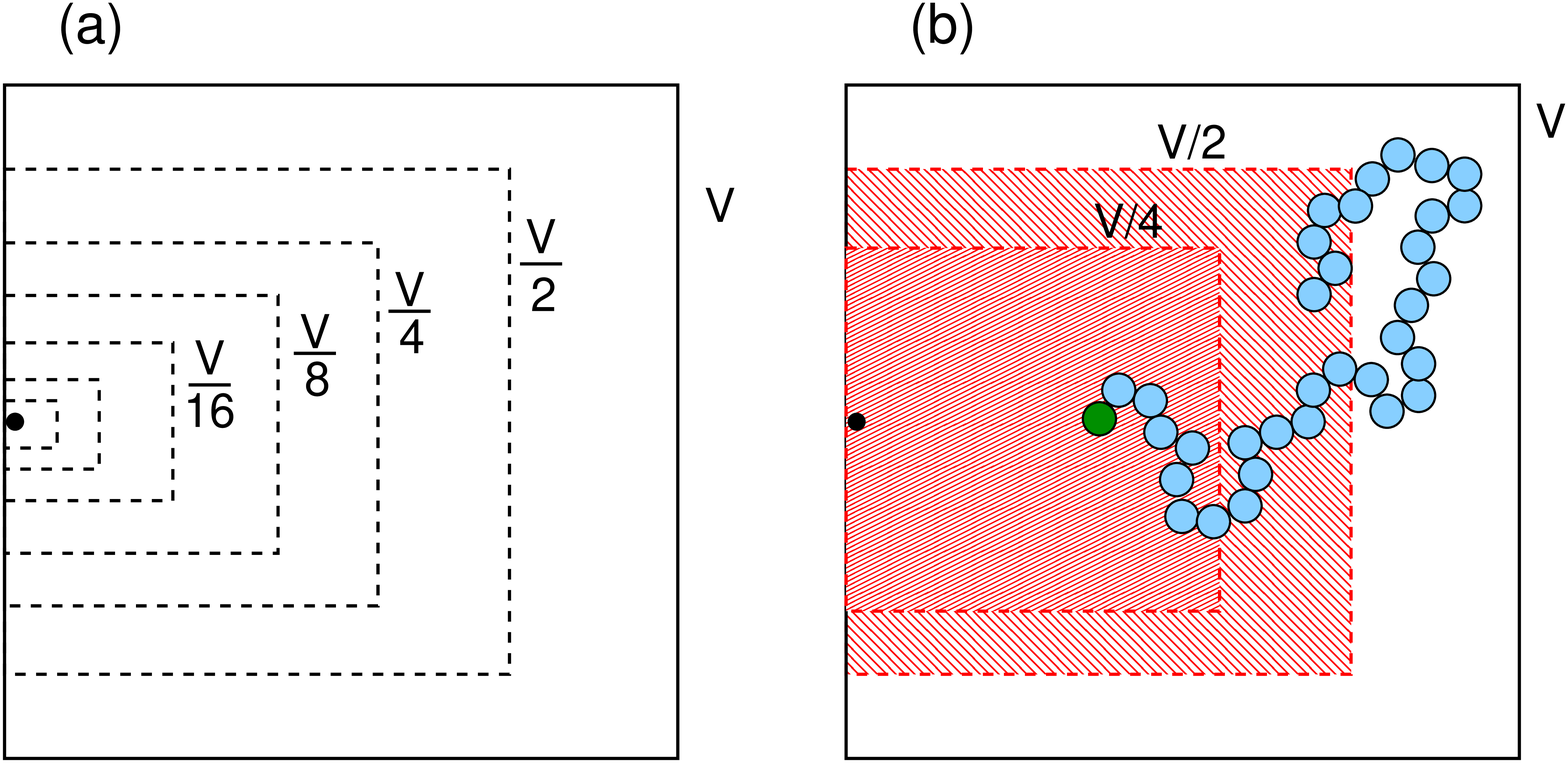}
\end{center}
\caption{Schematic illustration of the method used to calculate the chain-end localization 
free energy for volume subdivision fraction $\alpha_m=\frac{1}{2}$. 
(a) The volume subdivisions used to localize the chain end to a point (labeled 
as a black dot) on the inner surface of confinement. Each successive subvolume is half of 
the previous subvolume in the sequence.  (b) Illustration of method for calculating 
$\Delta F_m$ for $m$=2.  The selected end monomer (colored green) is confined to lie within
the volume $V/2$ (shaded red), while the remaining monomers (colored blue) can be anywhere 
in the full volume $V$.  $\Delta F_2$ is determined in a MC simulation by calculating the 
fraction of positions of the end monomer that lie within the volume $V/4$ (double-shaded red).
}
\label{fig:illust}
\end{figure}

To summarize the procedure formulated in this section, we can use the method described in 
Section~\ref{subsec:partial} to measure the difference in the conformational free energy 
$\Delta F_m$ in Eq.~(\ref{eq:FmV2_2}) for successive reductions in the end-monomer confinement 
volume. These values are used in Eq.~(\ref{eq:DFint2}) to estimate the change in the internal 
conformational free energy of the polymer upon localization of the end monomer to a 
target location inside the cavity.  As $n$ (the number of subdivisions of the end-monomer 
confinement volume) increases, the estimate for $\Delta F_{\rm loc}$ in Eq.~(\ref{eq:DFint2}) 
is expected to converge to the correct value.  Figure~\ref{fig:illust} provides a schematic 
illustration of the method used to calculate $\Delta F_m$ used in Eq.~(\ref{eq:FmV2}).

Note we have presented the development of the method assuming (1) end-monomer localization and 
(2) no other constraints (other than cavity confinement) are present. However, it is easily
shown that the method is equally valid for localizing a monomer located at arbitrary position
along the polymer contour or the case of monomer localization for a polymer with one end already 
tethered to a different point on the surface. Both of these other cases are also examined in this
study.

\hspace*{0.2in}
\section{Model}
\label{sec:model}

We model the polymer as a freely-jointed chain of hard spheres, each with
diameter $\sigma$.  The pair potential for non-bonded monomers is thus $u_{\rm{nb}}(r)=\infty$
for $r\leq\sigma$ and $u_{\rm{nb}}(r)=0$ for $r>\sigma$, where $r$ is the distance between
the centers of the monomers. Pairs of bonded monomers interact with a potential
$u_{\rm{b}}(r)= 0$ if $\sigma<r<1.15\sigma$ and $u_{\rm{b}}(r)= \infty$, otherwise.

The polymer is confined to a cavity that is either cubic or spherical in shape. The cavity
walls are ``hard,'' such that the monomer-wall potential energy is $u_{\rm wall}=\infty$
if the monomer center lies with a distance of $\sigma/2$ from the nearest point on the wall,
and $u_{\rm wall}=0$, otherwise. The width of the cavity $D$ is defined $D\equiv V^{1/3}$,
where $V$ is the volume of the box accessible to the monomer centers. Thus, $D=D_{\rm true}-\sigma$
for a cubic cavity with an actual width of $D_{\rm true}$, and $D=(\pi/6)^{1/3}(D_{\rm true}-\sigma)$
for a spherical cavity of actual diameter $D_{\rm true}$. 
The subvolumes for end-monomer confinement used in the localization algorithm were chosen to
be hard-walled cavities of the same shape as the true confinement cavity. Thus, cubic subvolumes
were used for cubic cavities and spherical subvolumes for spherical cavities.

\section{Simulation Details}
\label{sec:methods}

We employ the Metropolis Monte Carlo method to generate the states of the confined polymer
system, where a state is defined by the collection of monomer coordinates. In most of the
simulations we calculate the localization free energy of an end monomer, though we 
also consider the case of localizing a monomer at arbitrary position along the polymer.
Since the model system, including the artificial potentials imposed on the selected monomer, 
is athermal in character, trial moves are rejected if the move results in overlap of a 
monomer with another monomer or the wall, or if it violates the constraints associated
with the bonds or localized-monomer confinement; otherwise, it is accepted with 100\% probability.
Trial moves are generated using a combination of random monomer displacement moves, crankshaft 
moves, reptation, and sequence-inversion moves. A sequence-inversion move is defined by the 
monomer coordinate exchange ${\bf r}_{i}\rightarrow {\bf r}_{N-i}$. Such moves do not change
the shape or position of the polymer as a whole but do change the position of the monomer
to be localized. In the case of end-monomer localization, the large end-monomer displacements 
that are produced are accepted with reasonably high probability only for the larger subvolumes 
in the earlier stages of the calculation, i.e. for lower values of the index $m$ in Eq.~(\ref{eq:Vmm}). 
For smaller subvolumes, the acceptance ratio approaches zero and the moves are not used.
In some simulations we localize the end monomer of a polymer whose other end is already
tethered to a point on the confining surface. In those cases we cannot use reptation or
sequence inversion moves.

The number of volume subdivisions $n$ used to calculate the localization free energy was
chosen to be between $n$=12 and $n$=30, where large $n$ was used for larger cavity sizes.
The final subvolume was chosen to be $V_n/\sigma^3=0.02^3$, which fixes the values of the volume 
scaling constants $\alpha_m$ appearing in Eq.~(\ref{eq:Vmm}). For example, for a $D$=30 cubic 
cavity used for a $N$=100 polymer, using $n$=12 leads to to be $\alpha_m$=0.16069 for all $m$.

At each stage of the calculation for a given end-monomer subvolume, the polymer was first 
equilibrated, following which a production run was used to acquire the data. As an example, for 
a simulation with a $N$=200 polymer in a cubic cavity of width $D$=50, the system was equilibrated 
for $5\times 10^6$ MC cycles, following which a production of $1\times 10^8$ MC cycles was carried 
out. A single-monomer move (translation or crankshaft), a reptation move and an inversion move are 
each attempted on average once during each MC cycle. Maximum displacements for the single-monomer 
translational and crankshaft moves were chosen to yield acceptance ratios near 50\%. Each free 
energy calculation was performed between 5 and 10 times using different random number sequences. 
The statistically independent results were then used in the estimation of the standard error.

In the results presented below, distances are measured in units of $\sigma$ and energy 
is measured in units of $k_{\rm B}T$.

\section{Results}
\label{sec:results}

\subsection{End-monomer localization}
\label{subsec:endmon}

\begin{figure}[!ht]
\begin{center}
\vspace*{0.2in}
\includegraphics[width=0.45\textwidth]{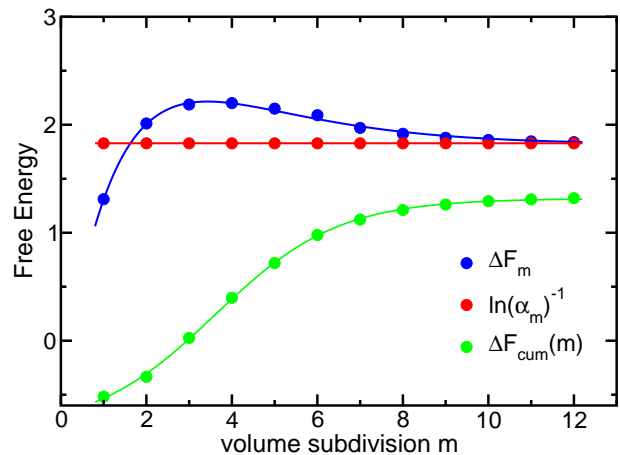}
\end{center}
\caption{Illustration of the convergence of the summation of Eq.~(\ref{eq:DFint2}) for
the case of confinement in a cubic cavity of side length $D$=30 and a polymer of
length $N$=100.  Here, the volume subdivision ratio is $\alpha_{m}=0.16069$ for the $m$th 
subdivision.  $\Delta F_m$ represents the free-energy difference between localizing the 
end-monomer volumes $V_m$ and $V_{m-1}$, where these subvolumes are defined in 
Eq.~(\ref{eq:Vmm}).  In addition, $\ln(\alpha_m^{-1})$ is the free-energy difference (in 
units of $k_{\rm B}T$) for a single isolated particle for confinement in these two subvolumes. 
These two quantities appear on the right side of Eq.~(\ref{eq:DFint2}). $\Delta F_{\rm cum}(m)$ 
is the cumulative sum of the difference in these two quantities. As the number of volume 
subdivisions $m$ increases, $\Delta F_{\rm cum}(m)$ converges to $\Delta F_{\rm loc}$. 
}
\label{fig:Fconv.cube.N100.L30}
\end{figure}

Figure~\ref{fig:Fconv.cube.N100.L30} presents simulation results that illustrate the convergence
of the summation of Eq.~(\ref{eq:DFint2}). For this case, the end monomer of a $N$=100 polymer 
is localized to a point in the middle of one face of a confining cube of dimension $D$=30.  
The graph shows 
$\Delta F_m$ vs $m$, where $\Delta F_m$ is the free-energy difference between localizing the
end-monomer volumes $V_m$ and $V_{m-1}$, where the subvolumes are defined in Eq.~(\ref{eq:Vmm}). 
In addition, the volume subdivision index ranges from $m$=1 to $m$=12. Also shown is the cumulative
summation $\Delta F_{\rm cum}(m)\equiv \sum_{i=1}^m (\Delta F_i - \ln\alpha_i^{-1})$.
As the number of volume subdivisions increases and thus the localization volume of the end monomer
becomes very small, $\Delta F_m$ converges to $\ln\alpha_m^{-1}$. This indicates that any
further reduction in the end-monomer confinement volume results simply in the reduction in
translational entropy of that monomer and not to any change to the conformational free energy
of the polymer. Thus, the value of the cumulative value $\Delta F_{\rm cum}(m)$ levels off to
the localization free energy, $\Delta F_{\rm loc}$.

\begin{figure}[!ht]
\begin{center}
\vspace*{0.2in}
\includegraphics[width=0.45\textwidth]{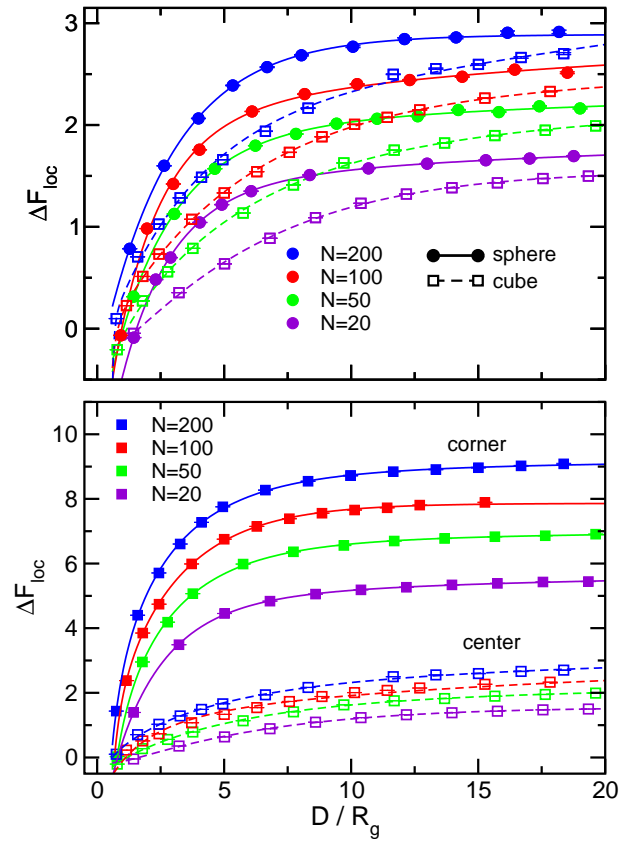}
\end{center}
\caption{(a) $\Delta F_{\rm loc}$ vs confining $D/R_{\sf g}$ for chains of length 
$N$=20, 50, 100 and 200. Results are shown for both spherical cavities (solid circles) and cubic 
confinement cavities (open squares). For the cubic cavities, chain end localization occurs at 
the center of one side of the cube.  Here, $D\equiv V^{1/3}$, where $V$ is the volume 
of the confining cavity accessible to the monomer centers, and $R_{\rm g}$ is the radius of 
gyration of an unconfined polymer. (b) As in panel (a), except results are shown for localization
of a chain end to the corner of a cube (solid squares) and the center of one face of
a cube (open squares).}
\label{fig:F.R.N.all}
\end{figure}

Figure~\ref{fig:F.R.N.all} shows the localization free energy $\Delta F_{\rm loc}$ as a function
of $D/R_{\rm g}$, the confinement cavity size scaled by the radius of gyration of a free polymer,
$R_{\rm g}$. Results for polymer lengths of $N$=20, 50, 100 and 200 are shown. Figure~\ref{fig:F.R.N.all}(a) 
shows free energies for end-monomer localization to a point on the surface of a confining sphere (solid
symbols and lines) and for localization the mid-point of one face for confinement in
a cubic cavity (open symbols and dashed lines).  Figure~\ref{fig:F.R.N.all}(b) compares two sets
of results for cubic confinement: one for localization to a mid-point on a face (open symbols and
dashed lines) and the other for localization to a corner of the cube (closed symbols and solid lines).
The curves for each data set are shown as guides for the eye and are not fits using any theoretical
prediction. In general, the free energy increases monotonically with cavity size and gradually levels
off to some asymptotic value. $\Delta F_{\rm loc}$ rises rapidly with $D/R_{\rm g}$ for 
$D/R_{\rm g}\lesssim 5$ and is nearly constant for $D/R_{\rm g}\gtrsim 10$. For $D/R_{\rm g}\lesssim 1$
$\Delta F_{\rm loc}$ is even slightly negative. Thus, for a very tightly confined polymer, the end 
monomer has a higher probability of being at the point on a wall than would be the case for the monomer 
in the absence of the remaining portion of the polymer. We also note that $\Delta F_{\rm loc}$ is 
generally higher for spherical confinement compared to cubic confinement (with mid-face localization), 
though the two sets of results appear to converge to the same value as the confinement volume becomes 
very large.  On the other hand, in the case of cubic confinement, the free energy for localization 
to a corner is consistently larger than that for the mid-face localization, even in the limit of 
very large confinement volumes. Finally, we note that in all cases $\Delta F_{\rm loc}$ increases 
monotonically with polymer length.

\begin{figure}[!ht]
\begin{center}
\vspace*{0.2in}
\includegraphics[width=0.45\textwidth]{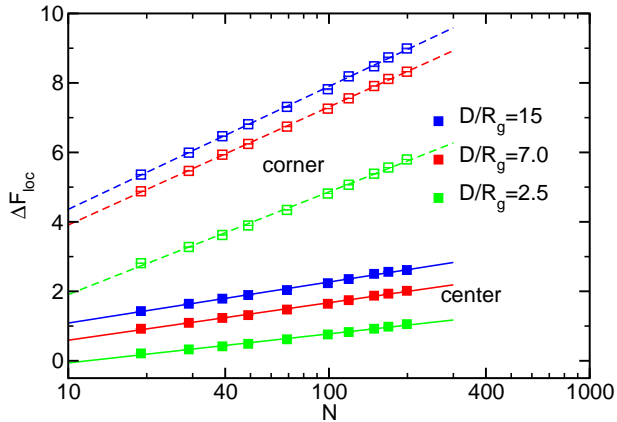}
\end{center}
\caption{(a) $\Delta F_{\rm loc}$ vs polymer length $N$ for polymers confined to a cubic space. 
Results are shown for various ratios of $D/R_{\rm g}$ for the cases of chain-end 
localization to the center of one face of the cube (closed squares) and to one corner of the cube 
(open squares). The curves overlaid on the data are fits to $\Delta F_{\rm loc} = c + \alpha\ln N$.
The fitting parameter values are as follows for localization to a corner: 
$\alpha$=$1.54\pm 0.01$ and $c$=$0.82\pm 0.06$ for $D/R_{\rm g}$=15;
$\alpha$=$1.48\pm 0.01$ and $c$=$0.51\pm 0.05$ for $D/R_{\rm g}$=7.5;
$\alpha$=$1.28\pm 0.02$ and $c$=$-1.04\pm 0.08$ for $D/R_{\rm g}$=2.5.
For localization to the center of a face, the values are:
$\alpha$=$0.51\pm 0.01$ and $c$=$-0.09\pm 0.04$ for $D/R_{\rm g}$=15;
$\alpha$=$0.47\pm 0.01$ and $c$=$-0.49\pm 0.04$ for $D/R_{\rm g}$=7.5;
$\alpha$=$0.36\pm 0.01$ and $c$=$-0.90\pm 0.04$ for $D/R_{\rm g}$=2.5.
}
\label{fig:F.N.ratio.cube}
\end{figure}

Figure~\ref{fig:F.N.ratio.cube} shows the variation of $\Delta F_{\rm loc}$ with polymer length
for three different cavity sizes. In each case results are shown for chain localization at the 
center of one face of the confining cube and at one corner of the sphere. As is evident in the 
figure, $\Delta F_{\rm loc}$ varies approximately linearly with $\ln N$. In addition, 
$\Delta F_{\rm loc}$ increases more rapidly with polymer length for end-monomer localization
at a corner than for the case of localization near a cube face center. Consistent with the results
of Fig.~\ref{fig:F.R.N.all} the free energy increases with increasing confinement volume in all 
cases.

Each of the trends evident in Figs.~\ref{fig:F.R.N.all} and \ref{fig:F.N.ratio.cube} can be 
accounted for using standard scaling arguments together with simple theoretical modeling. 
Let us first consider the origin of the scaling of $\Delta F_{\rm loc}$ with $N$ for the case 
of a very weakly confined polymer where $D/R_{\rm g}\gg 1$. It is useful to first note
that in a scale-free environment the conformational contribution to the partition function of 
a self-avoiding chain can be written
\be
Z_i = q^{N} N^{\gamma_i-1}
\label{eq:Zi}
\ee
where $N$ is the number of bonds of a polymer composed of $N+1$ monomers and $q=e^{-\mu/k_{\rm B}T}$,
where $\mu$ is the chemical potential per monomer of the chain. In addition, $\gamma_i$ is a
critical exponent whose value depends on whether the polymer is free or tethered to a surface.
For a free polymer ($i$=a), $\gamma_a\approx 1.16$, while for a polymer tethered to an infinite 
flat surface ($i$=b), $\gamma_{\rm b}\approx 0.69$.\cite{Muthukumar_book} Since the free energy 
of a polymer is given by $F_i/k_{\rm B}T=-\ln Z_i$, it follows that the conformational free-energy 
difference between a free polymer and one tethered to a surface is 
$\Delta F_{\rm ab} \equiv F_{\rm b}-F_{\rm a} = \alpha_{\rm ab}\ln N$, where 
$\alpha_{\rm ab}\equiv \gamma_{\rm a} -\gamma_{\rm b} \approx 0.47$. Clearly, $\Delta F_{\rm ab}$ 
is equal to $\Delta F_{\rm loc}$ for chain-end localization to the center of a cubic box in the 
limit of infinite box size.  For the largest confining cube we consider of $D/R_{\rm g}$=15, 
a fit to the data yields $\alpha=0.51\pm 0.01$, which is comparable to though slightly larger than
the theoretical prediction.

Now consider the case of a polymer tethered to the corner of an semi-infinite octant of space 
(e.g. the polymer is tethered at the origin and the polymer is constrained to lie in the 
region $x>0$, $y>0$ and $z>0$). We label the index in Eq.~(\ref{eq:Zi}) $i$=c for this case.
The conformational free-energy difference between this tethered polymer and a free polymer is 
$\Delta F_{\rm ac}\equiv F_{\rm c}-F_{\rm a} = \alpha_{\rm ac} \ln N$, where 
$\alpha_{\rm ac}\equiv \gamma_{\rm c}-\gamma_{\rm a}$.
This is the predicted form of $\Delta F_{\rm loc}$ for end-monomer localization at the corner
of a confining cube in the limit of infinite cube size. For the largest confining cube we
use (i.e. $D/R_{\rm g}$=15), a fit to the data yields  $\alpha=1.51\pm 0.02$.
We are unaware of whether the scaling exponent $\gamma_{\rm c}$ has been calculated for such a 
corner-tethered polymer. Consequently, we have used a different Monte Carlo simulation method
to measure the free energies $F_{\rm a}$, $F_{\rm b}$ and $F_{\rm c}$. The results 
are presented in Appendix~\ref{app:a}. We find that the length-dependence of the polymer 
satisfies the relation
\begin{eqnarray}
\Delta F_{\lambda\mu} = C_{\lambda\mu} + \alpha_{\lambda\mu}\ln N
\end{eqnarray}
%
%
The quantity of interest here was found to be $\alpha_{\rm ac}$=$1.451\pm 0.001$, which compares 
well to the value for the data for the largest confining cube in Fig.~\ref{fig:F.N.ratio.cube}.

The arguments above hold for large box size, i.e. $D/R_{\rm g}\gg 1$. In order to account 
for the variation of $\Delta F_{\rm loc}$ with box size, i.e. the localization free energy 
decreases with decreasing $D$, we develop a simple theoretical model based on the following 
simplifications.  If the polymer end lies in a layer of width $\approx R_{\rm g}$ near the walls 
of the box, the polymer will interact with the box, leading to an increase in the conformational 
free energy. If the end monomer lies a distance $\gtrsim R_{\rm g}$ away from the walls of the box,
the  polymer does not interact with the walls and its conformational free energy, $F_{\rm a}$, is 
otherwise independent of position. Figure~\ref{fig:F.scale} in Appendix~\ref{app:b} shows simulation 
results for the variation of the free energy with end-monomer position with distance from
a wall for a variety of chain lengths and confirms this expected trend. For simplicity, we model 
the surface effects on the free energy by choosing a constant value when the chain end lies in 
a region close to the confining surface. The value of this free energy is chosen to
be that calculated in Appendix~\ref{app:a} for a polymer tethered to a wall or corner.
The theoretical model is developed in Appendix~\ref{app:c}. The localization free energy is 
calculated using Eqs.~(\ref{eq:Fint_app})--(\ref{eq:Vabc}) for localization to the mid-point on 
a cubic face, Eqs.~(\ref{eq:DFmn}), (\ref{eq:Vabc}) and (\ref{eq:DFintcor}) for corner 
localization inside a cube, and Eqs.~(\ref{eq:DFmn}), (\ref{eq:DFintsph}) and (\ref{eq:Vab}) 
for localization to a point on the inside wall of a spherical cavity.

\begin{figure}[!ht]
\begin{center}
\vspace*{0.2in}
\includegraphics[width=0.45\textwidth]{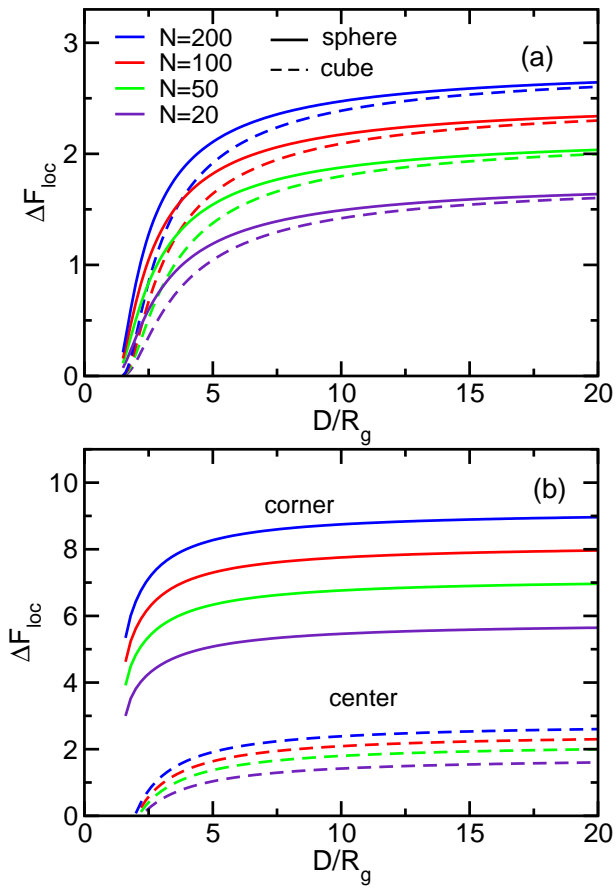}
\end{center}
\caption{Predictions for $\Delta F_{\rm loc}$ using the theoretical model of Appendix~\ref{app:c}.
(a) $\Delta F_{\rm loc}$ vs $D/R_{\rm g}$ for $N$=20, 50, 100 and 200. In each case results are 
shown for end-monomer localization to  the center of one face of the cubic box and to a
point on the wall of a confining sphere. (b) As in panel (a), except results are shown for end-monomer 
localization to  the center of one face of the cubic box and localization to one corner
of the cube.  }
\label{fig:theory.cube.sphere}
\end{figure}

\begin{figure}[!ht]
\begin{center}
\vspace*{0.2in}
\includegraphics[width=0.45\textwidth]{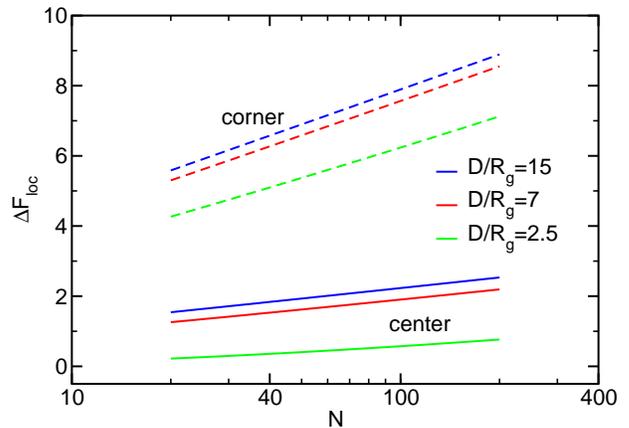}
\end{center}
\caption{Predictions for $\Delta F_{\rm loc}$ using the theoretical model of Appendix~\ref{app:c}.
$\Delta F_{\rm loc}$ vs $N$ for $D/R_{\rm g}$=2.5, 7 and 15. In each case results are
shown for end-monomer localization to  the center of one face of the cubic box and localization 
to one corner of the cube.}
\label{fig:theory.N}
\end{figure}

Figure~\ref{fig:theory.cube.sphere}(a) shows the prediction for the variation of $\Delta F_{\rm loc}$ 
with box size for several different polymer lengths. Results are shown for spherical and cubic box 
confinement with end-monomer localization to the center of one face of the confining cube and to 
one point on the confining sphere.  Likewise, Fig.~\ref{fig:theory.cube.sphere}(b) compares results 
for $\Delta F_{\rm loc}$ for cubic confinement with end-monomer localization to the center of one 
face of the confining cube and localization to one of its corners.  Figure~\ref{fig:theory.N} shows 
predicted variation of $\Delta F_{\rm loc}$ with polymer length for three different cubic box sizes. 
Results are shown for localization to both the center of one face and to one corner of the confining 
cube. The model correctly predicts the main qualitative trends including: (1) the rapid decrease in 
$\Delta F_{\rm loc}$ with decreasing cavity size for $D/R_{\rm g}\lesssim 4$; (2) the higher value
of the free energy for localization to a corner vs the center of a face of the confining cube; (3)
the consistently higher value of $\Delta F_{\rm loc}$ for confinement in a sphere compared to
a cube of the same volume (with face-centered tethering); (4) the linear variation of 
$\Delta F_{\rm loc}$ with $\ln N$ with approximately the same scaling factor. The quantitative
discrepancies naturally arise from the approximations we have employed, notably choosing
the conformational free energy to be constant within a finite range of end-monomer distance from
the confining surface. However, the semi-quantitative agreement demonstrates that our basic
interpretation of the trends is correct.

\subsection{Localization of arbitrary monomer}
\label{subsec:intmon}

\begin{figure}[!ht]
\begin{center}
\vspace*{0.2in}
\includegraphics[width=0.45\textwidth]{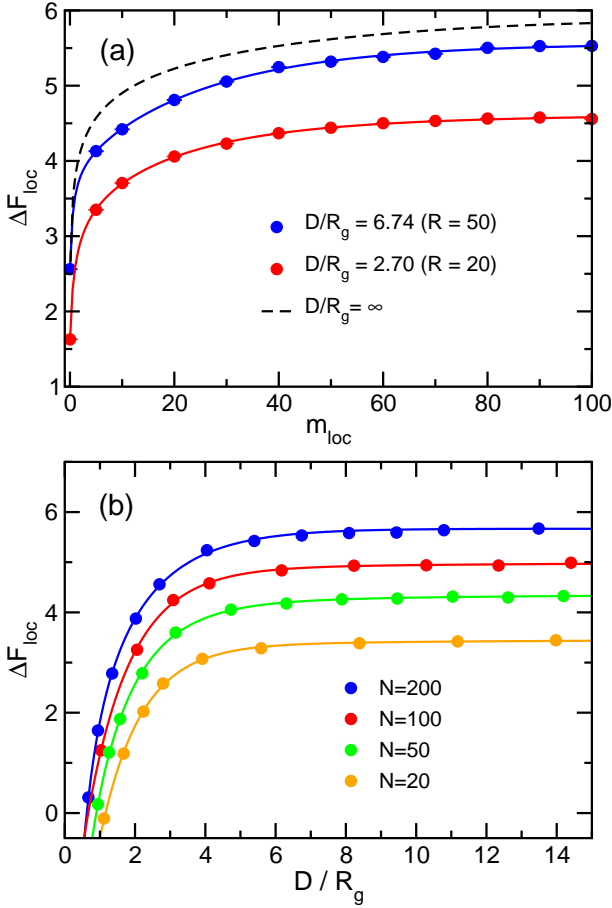}
\end{center}
\caption{(a) $\Delta F_{\rm loc}$ vs scaled confining cavity size $D/R_{\sf g}$, where 
$\Delta F_{\rm loc}$
is the free-energy cost of localizing a middle monomer to a point in the center of a face
on a confining cubic surface. Results are shown for polymer chains of length $N$=20, 50, 100 
and 200. The dashed line shows the theoretical prediction of Eq.~(\ref{eq:Fmloc_th}) for the 
limiting case of $D/R_{\rm g}\rightarrow\infty$. The constant in Eq.~(\ref{eq:Fmloc_th}) is 
chosen simply to shift the theoretical curve above the simulation data for clarity.
(b) $\Delta F_{\rm loc}$ vs monomer localization index $m_{\rm loc}$ for a $N$=200
polymer confined to a spherical cavity. Results for different spherical cavity sizes are shown.}
\label{fig:delF.D.middle}
\end{figure}

Thus far we have examined the free energy of localizing a single end monomer to a point
on the inside wall of a confining cavity. Consider now the case of localization of a 
monomer at arbitrary position along the contour of the polymer. We denote the index for this
monomer $m_{\rm loc}$, which ranges from 0 to $N$ for a polymer of $N+1$ monomers.
Figure~\ref{fig:delF.D.middle}(a) shows the variation of $\Delta F_{\rm loc}$ with 
$m_{\rm loc}$ for a polymer of length $N$=200. Note that in the figure $m_{\rm loc}$ is varied 
from from the position of one end monomer ($m_{\rm loc}=0$) to the middle monomer 
($m_{\rm loc}=100$); the free energy for the range $m_{\rm loc}=100$ to 200 is related by
symmetry. Results are shown for two different cavity sizes. In each case
$\Delta F_{\rm loc}$ is smallest at the end position and increases rapidly with $m_{\rm loc}$
until about $m_{\rm loc}\approx 20$, after which it increases at a much slower rate. The
free-energy maximum lies at $m_{\rm loc}=N/2=100$, i.e. the case where the localized monomer
is at the midpoint along the contour of the polymer. At each value of $m_{\rm loc}$, the
free energy increases with increasing cavity size.
Figure~\ref{fig:delF.D.middle}(b) shows the variation of $\Delta F_{\rm loc}$ with cavity size in
the case of localization of the central monomer, i.e. $m_{\rm loc}=N/2$. Results are shown for
different polymer lengths. The general trends are qualitatively consistent with the results
in Fig.~\ref{fig:F.R.N.all} for end-monomer localization; that is, $\Delta F_{\rm loc}$ increases
rapidly with $D$ for $D/R_{\rm g}\lesssim 3$ and gradually levels off for larger $D$, and it
increases with increasing $N$. The origins of these effects are the same as those discussed
for end-monomer localization.

\begin{figure}[!ht]
\begin{center}
\vspace*{0.2in}
\includegraphics[width=0.45\textwidth]{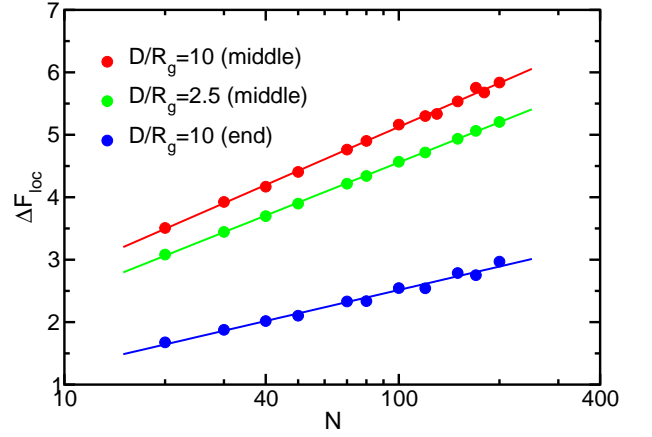}
\end{center}
\caption{$\Delta F_{\rm loc}$ vs polymer length $N$ for a polymer confined to a sphere in the case of
localization of the middle monomer of the polymer.  Results are shown for $D/R_{\rm g}$=2.5 and 10.
For convenient comparison, results are also shown for end-monomer localization with $D/R_{\rm g}$=10.
The curves overlaid on the data are fits to $\Delta F_{\rm loc} = c + \alpha\ln N$.
The fitting parameter values are as follows for localization of the middle monomer:
$\alpha$=$1.01\pm 0.02$ and $c$=$0.46\pm 0.07$ for $D/R_{\rm g}$=10, and
$\alpha$=$0.928\pm 0.005$ and $c$=$0.28\pm 0.02$ for $D/R_{\rm g}$=2.5.
For end-monomer localization with $D/R_{\rm g}$=10, $\alpha$=$0.54\pm 0.02$ and $c$=$0.0\pm 0.1$. 
}
\label{fig:delF.N.midpoint}
\end{figure}

Figure~\ref{fig:delF.N.midpoint} shows the variation of the free energy with chain length in
the case of $m_{\rm loc}=N/2$ for a polymer confined to a spherical cavity. Results are shown
for cavity sizes of $D/R_{\rm g}$=2.5 and 10. For a helpful comparison, the free energy for a polymer 
with end-monomer localization ($m_{\rm loc}$=0) is also shown.  We find that $\Delta F_{\rm loc}$ 
varies linearly with $\ln N$, as was the case in Fig.~\ref{fig:F.N.ratio.cube} for end-monomer 
localization inside a cube. The solid curves in the figure are fits to the function
$\Delta F_{\rm loc} = c + \alpha\ln N$.  For $m_{\rm loc}=N/2$, we find that $\alpha=1.01\pm 0.02$
for $D/R_{\rm g}$=10 and $\alpha=0.928\pm 0.05$ for $D/R_{\rm g}$=2.5. By contrast, we
find that $\alpha=0.54\pm 0.02$ for $m_{\rm loc}$=0 and $D/R_{\rm g}$=10.

An explanation for the variation in $\Delta F_{\rm loc}$ with localization position
is given as follows. Consider first the case of a very large confining cavity, in which case
the confining surface near the localization point is effectively flat.  Next, note 
that a polymer with monomer $m_{\rm loc}$ localized to a point on the wall corresponds to two 
subchains of length $m_{\rm loc}$ and $N-m_{\rm loc}$, each connected to this point.  
As noted earlier, the partition function for a single self-avoiding 
chain tethered to a point on an infinite flat wall is $Z_1(N)=q^N N^{\gamma_1-1}$, where 
$\gamma_1\approx$ 0.69 and where $q$ is the effective coordination number of the random walk.  
Thus, in the hypothetical case where the two subchains interact internally but not with each 
other, the partition function is $Z(m_{\rm loc})=Z_1(m_{\rm loc})Z_1(N-m_{\rm loc})$, or
\begin{eqnarray}
Z(m_{\rm loc}) = q^{N} \left(m_{\rm loc}(N-m_{\rm loc}\right))^{\gamma_1-1}.
\label{eq:Zmloc1}
\end{eqnarray}
For localization at the mid-point of $m_{\rm loc}=N/2$, this becomes 
$Z(m_{\rm loc})= q^N (N/2)^{2\gamma_1-1}$. As noted in Ref.~\onlinecite{gaunt1990branched}, 
interactions between the subchains leads to a modified scaling of $Z\sim q^{N} (N/2)^{\gamma_2-1}$ 
where $\gamma_2\approx 0.203$.  Thus, for $m_{\rm loc}=N/2$ we find that 
$F/k_{\rm B}T = -\ln Z = (1-\gamma_2)\ln (N) -N\ln q +$const.  
Since the partition function of a free self-avoiding polymer scales as 
$Z_0\sim q^N N^{\gamma_0-1}$, where $\gamma_0\approx$1.16, it follows that the difference in 
the conformational free energy for a free polymer and one with the middle monomer localized 
to a point on a flat surface is given by
\begin{eqnarray}
\Delta F/k_{\rm B}T = (\gamma_2-\gamma_1)\ln N + {\rm const.} 
\end{eqnarray}
where $\gamma_2-\gamma_0$=0.96. This scaling is expected for the free 
energy cost of localizing a middle monomer in the limit of infinitely large confinement
cavity and for a sufficiently long chain. This scaling factor is close to the value of 
$\alpha = 1.01\pm 0.02$ observed for the large cavity size of $D/R_{\rm g}=10$.
As noted earlier, in the case of end-monomer localization, the corresponding prediction
for the localization free energy is $\Delta F = 0.47\ln N$. This is comparable to the
observed scaling of $\Delta F_{\rm loc}=(0.54\pm 0.02)\ln N$ +constant observed for 
end-monomer localization in a spherical cavity of size $D/R_{\rm g}=10$. As before, the 
small differences between the predicted and observed scaling are attributable to finite-size
effects associated with the finite cavity size and the polymer length.

To account for the variation of $\Delta F_{\rm loc}$ with $m_{\rm loc}$ observed in 
Fig.~\ref{fig:delF.D.middle}(a), we follow Ref.~\onlinecite{mihovilovic2013statistics} and
note that application of Duplantier's theory of polymer networks\cite{duplantier1989statistical} 
leads to a modification of Eq.~(\ref{eq:Zmloc1}) to 
$Z(m_{\rm loc})\sim q^N m_{\rm loc}^{\gamma_2-\gamma_1} (N-m_{\rm loc})^{\gamma_1-1}$.
Consequently, the localization free energy, $\Delta F_{\rm loc}(m_{\rm loc})/k_{\rm B}T=-\ln(Z/Z_0)$,
can be written as
\begin{eqnarray}
\Delta F_{\rm loc}/k_{\rm B}T & = & 0.49\ln m_{\rm loc} + 0.31\ln(N-m_{\rm loc}) + c(N), \nn\\
\label{eq:Fmloc_th}
\end{eqnarray}
where $c(N)=0.16\ln N + c_0$ and where $c_0$ is an unknown constant. The theoretical prediction 
for $\Delta F_{\rm loc}(m_{\rm loc})$ is plotted in Fig.~\ref{fig:delF.D.middle} using a constant 
of $c_0=1.3$ to shift the curve above the simulation data for clarity. The predicted function is 
comparable to the measured $\Delta F_{\rm loc}(m_{\rm loc})$.

\subsection{End-monomer localization of a tethered polymer}
\label{subsec:tether}

\begin{figure}[!ht]
\begin{center}
\vspace*{0.2in}
\includegraphics[width=0.45\textwidth]{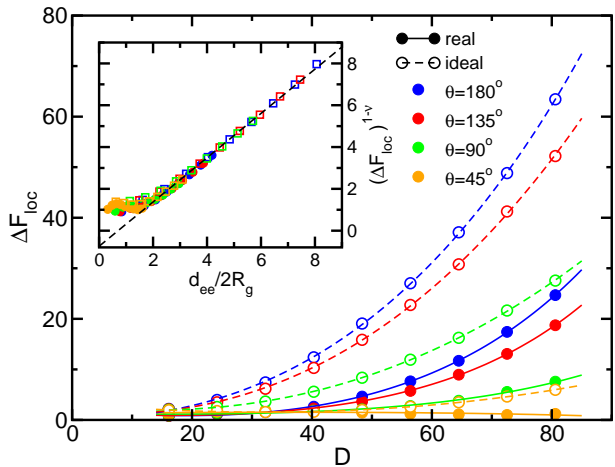}
\end{center}
\caption{$\Delta F_{\rm loc}$ vs confining sphere diameter $D$ for a chain of length $N$=200. 
Here, $\Delta F_{\rm loc}$ is the change in the conformational free energy upon localizing a 
monomer to a point on the inner surface of the sphere in the case where the other monomer is 
tethered to a different point on the sphere. Results are shown for four different values of 
the angular distance $\theta$ between the tethered point and the localization point for 
real (i.e. self-avoiding) polymers and ideal polymers. The inset shows the scaled free energy 
$(\Delta F_{\rm loc})^{1/(1-\nu)}$ vs scaled end-to-end distance $d_{\rm ee}/2R_{\rm g}$ where 
$d_{\rm ee} \equiv (6/\pi)^{1/3}D\sin(\theta/2)$. The results were calculated using the raw 
data with $\nu=\frac{1}{2}$ for ideal chains and $\nu=\frac{3}{5}$ for real chains.The 
dashed curve is a linear fit to the data in the domain $d_{\rm ee}/2R_{\rm g}>1.7$.} 
\label{fig:delF.sphere.tether.scale}
\end{figure}

Let us now consider the case of localization of an end monomer for confinement inside a sphere
in the case where the other end monomer is already tethered to a point on the surface.
Figure~\ref{fig:delF.sphere.tether.scale} shows $\Delta F_{\rm loc}$ vs cavity size $D$
for a polymer of length $N$=200. Results are shown for both real and ideal polymers and
for four values of the angle $\theta$ between the tethering point and the localization point
on the sphere. Several trends are evident. As expected the free energy increases monotonically
with increasing sphere size, due mainly to the stretching of the polymer. In addition, at fixed
$D$ the free energy increases with increasing $\theta$. This results from the increase in
the distance between tethering points with $\theta$ leading to further stretching of the polymer.
The effect of including monomer-monomer repulsion leads to a decrease in the free energy
relative to the case for ideal polymers. This arises because such repulsions lead to a swelling
of the chain leading to a tendency for the free end monomer to lie on average further away from the 
tethered end monomer and thus closer to the point where the free monomer is localized. 
Concomitantly, the probability that it lies at that point is greater and thus the localization
free energy is lower.

Another noteworthy feature of the data in Fig.~\ref{fig:delF.sphere.tether.scale} is the 
fact that localization free energy levels off for real chains in the range $D\lesssim 30$ and
appears to approach a minimum for ideal chains at $D\approx 10$. The significance
of this trend can be understood by noting that the effective diameter of the unconfined
polymer is $2R_{\rm g}=23.9$ for a real polymer and $2R_{\rm g}=12.4$ for an ideal polymer.
In each case, it is expected that the free-energy cost of pulling the ends of a polymer
a distance $\lesssim 2R_{\rm g}$ will be roughly independent distance and of the order of 
$k_{\rm B}T$ and independent of distance. Only when the distance exceeds this value will 
the polymer become appreciably distorted in shape. Consequently, the free energy is expected 
to rise considerably only after that point.

In order to carry out a quantitative analysis of the data we employ a standard expression for 
the scaling of the free energy of a stretched polymer of length $N$ with distance $d_{\rm ee}$ 
between the ends of the polymer:\cite{Rubenstein_book} 
$\Delta F/k_{\rm B}T \sim (d_{\rm ee}/R_{\rm g})^{1/(1-\nu)}$. Note that this relation employs 
the de~Gennes blob picture and is only expected to be valid when $d_{\rm ee}$ exceeds the 
average size of the free polymer, for which we choose $2R_{\rm g}$ as a convenient measure.  
Consequently, $(\Delta F/k_{\rm B}T)^{1-\nu}$ will scale linearly with $d_{\rm ee}$
only for $d\gtrsim 2R_{\rm g}$. In applying this result to the present analysis we initially
assume that confinement effects are weak and also note that the distance $d_{\rm ee}$ scales with 
$\theta$ as $d_{\rm ee}=(6/\pi)^{1/3} D\sin(\theta/2)$, where the factor of $(6/\pi)^{1/3}$ arises 
from the definition of $D\equiv V^{1/3}$. This leads to the prediction that 
$(\Delta F_{\rm loc})^{1-\nu}$ varies linearly with $d_{\rm ee}/2R_{\rm g}$ and, further, that 
the data collapses to a single curve for all values of $\theta$ and for real and ideal polymers. 
The inset of Fig.~\ref{fig:delF.sphere.tether.scale} shows that this prediction is borne out
since the data collapses to a single linear curve for $d_{\rm ee}/2R_{\rm g} \gtrsim 1.7$
This result is somewhat surprising for two reasons. First, it suggests that the confinement 
effects omitted from this analysis are not significant relative to the effects of stretching the 
polymer, at least in the regime where the end-to-end distance exceeds the mean size of the free 
polymer. In addition, for the larger values of $\theta$ and sphere size $D$, the end-to-end 
stretch distance is an appreciable fraction of the contour length. This leads to small blob 
sizes and eventually a breakdown in the whole blob picture that underlies the scaling relation 
we employ. 

\begin{figure}[!ht]
\begin{center}
\vspace*{0.2in}
\includegraphics[width=0.45\textwidth]{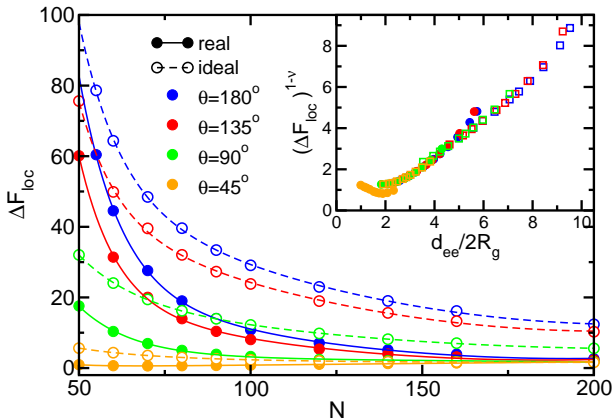}
\end{center}
\caption{As in Fig.~\ref{fig:delF.sphere.tether.scale}, except $\Delta F_{\rm loc}$ vs 
polymer length $N$ for confinement in a sphere of diameter $D$=50.  The inset shows 
$(\Delta F_{\rm loc})^{1-\nu}$ vs $d_{\rm ee}/2R_{\rm g}$, calculated as for the data 
in the inset of Fig.~\ref{fig:delF.sphere.tether.scale}.}
\label{fig:delF.sphere.N.tether.D50}
\end{figure}

Figure~\ref{fig:delF.sphere.N.tether.D50} shows the variation of $\Delta F_{\rm loc}$ with 
polymer length $N$ for a fixed sphere diameter of $D$=50. We consider chain lengths in the 
range of $N$=50--200. As in Fig.~\ref{fig:delF.sphere.tether.scale}, results are shown for various 
values of $\theta$ and for both real and ideal chains. As expected, the localization free 
energy cost increases as the chain length decreases and, thus,  as the relative 
degree of deformation of the polymer from its undistorted shape increases. Consistent 
with the previous results and for the same reasons outlined above, $\Delta F_{\rm loc}$ 
increases with increasing $\theta$ and is lower for real chains than for ideal chains. 
The inset shows $(\Delta F_{\rm loc})^{1-\nu}$ vs $d_{\rm ee}/2R_{\rm g}$. By comparison 
with Fig.~\ref{fig:delF.sphere.tether.scale} we note that the scaling of the data yields 
poorer collapse of the data and a less linear variation of $(\Delta F_{\rm loc})^{1-\nu}$ 
with the end-to-end distance. The deviations from such trends are especially evident for
the data points corresponding to large $d_{\rm ee}$, which here corresponds to the shortest
chain lengths. We believe that this is mainly due to a finite-size effect due to small $N$
and propose that better data collapse and more linear functions would emerge using polymer
lengths much greater than we can feasibly examine at present.

\section{Conclusions}
\label{sec:conclusions}

In this study we have presented and tested a Monte Carlo simulation method for calculating 
the free-energy cost of localizing a single monomer of a confined polymer to a point on the 
inner surface of the confining cavity. The core element of this method involves
measuring the probability that the chosen monomer lies in a volume $V_{\rm b}$, which is
a subvolume of a larger volume $V_{\rm a}$ within which the monomer is confined in a
given simulation. The volume $V_{\rm a}$ is itself a subvolume of the cavity volume $V$.
By carrying out a sequence of simulations in which $V_{\rm a}$ is varied from $V$ to
a very small volume situated at the localization point, summing the resulting free-energy 
differences, and subtracting out the translational free energy of the monomer, the localization
free energy $\Delta F_{\rm loc}$ is calculated.

Using this approach,  we measured $\Delta F_{\rm loc}$ for localization of an end monomer to 
sites on spherical and square confining surfaces and, in the latter case, to sites at the center 
of a cube face and at a cube corner. For a strongly confined polymer with $D\lesssim 2R_{\rm g}$
$\Delta F_{\rm loc}$ is negligible. As the confinement size $D$ increases,  $\Delta F_{\rm loc}$
initially increases rapidly and then levels off asymptotically in the weak confinement limit of
$D\gg R_{\rm g}$. The free-energy cost is significantly higher for end-monomer localization to
the corner of a cube than for localization to a cube face center, as expected from the greater
deformation of polymer chain that results for the former case. In all cases, $\Delta F_{\rm loc}$
increases monotonically with chain length $N$ and generally scales as 
$\Delta F_{\rm loc} = m\ln N$+constant.  At sufficiently large $D/R_{\rm g}$ the proportionality 
constant $m$ is consistent with the prediction that uses the known form of the partition function
of a self-avoiding chain in a scale-free environment,$Z\sim q^N N^{1-\gamma}$, along with the known
or measured values of the exponent $\gamma$ for a free polymer and a polymer end-tethered to a
flat wall or corner. In addition, a simple theoretical model yielded scaling results of 
$\Delta F_{\rm loc}$ with respect to $N$ and $D$ that are semi-quantitatively consistent with 
the simulation results. We also examined the case of localization of a monomer at arbitrary
position along the polymer contour and obtained results that are qualitatively similar to
those for end-monomer localization and quantitatively consistent with theoretical predictions.
This consistency provides a clear demonstration of the validity and accuracy of the method.
Finally, we examined end-monomer localization in the case where the other polymer end 
is tethered to a different point on the confinement surface. We find that the variation of 
$\Delta F_{\rm loc}$ with $D$ for a sufficiently long polymer is consistent with the standard
theoretical predictions for an unconfined stretched polymer using the blob model. 
This suggests that the effect of confinement in this case is not significant, at least for
tethering point distances that exceed the average polymer size of $\sim 2R_{\rm g}$.

All calculations in this work used a freely-jointed hard-sphere chain confined to a spherical 
or cubic cavity. Employing such a simple model was useful for demonstrating the validity and
utility of the method and for facilitating easy comparison with theoretical predictions.
In future work, it will be of interest to examine more complex models and consider the effects 
of features such as polymer bending rigidity, macromolecular crowding, and polymer topology (e.g.
ring or branched). It will also be useful to examine monomer localization for polymers
confined to cavities of a variety of types, including anisometrically shaped 
cavities\cite{polson2015polymer} and pyramidal cavities.\cite{langecker2011electrophoretic}
Finally, these calculations can be used to test directly those theories
of polymer translocation that emphasize the importance of the free-energy barrier on the 
translocation dynamics as the localization free energy can be calculated for any model used
in dynamics simulations.

\begin{acknowledgments}
This work was supported by the Natural Sciences and Engineering Research Council of Canada (NSERC).  
We are grateful to Compute Canada and the Atlantic Computational Excellence Network (ACEnet) for 
use of their computational resources.
\end{acknowledgments}

\appendix
\section{Free energy of free and tethered polymers}
\label{app:a}

In this appendix, we calculate the conformational free energies for (a) a free polymer, $F_{\rm a}$, 
(b) a polymer tethered to an infinitely wide flat hard wall, $F_{\rm b}$, and (c) a polymer tethered
to a corner of an infinitely large hard-walled cube, $F_{\rm c}$. These are used to obtain
the free-energy differences $\Delta F_{\rm ab}\equiv F_{\rm b}-F_{\rm a}$, 
$\Delta F_{\rm ac}\equiv F_{\rm c}-F_{\rm a}$ and $\Delta F_{\rm bc}=F_{\rm c}-F_{\rm b}$, which
are used in the model developed in Appendix~\ref{app:c} to estimate the chain-end localization
free energy for a polymer under confinement in a cavity of arbitrary size. In the limit of
large $N$ it is known that the partition function for such scale-free polymer systems has 
the form $Z_i = q^{N}N^{\gamma_i-1}$, where the effective coordination
number $q$ is related to the chemical potential per monomer $\mu$ by $q=e^{-\beta\mu}$.
In addition, the scaling exponent $\gamma_i$ depends on the constraints, if any, imposed on the polymer.
For a free polymer ($i$=a), $\gamma_{\rm a}=1.16$, while for a polymer tethered to wall ($i$=b), 
$\gamma_{\rm b}$=0.69. Noting that $\beta F_i = -\ln Z_i$, it follows that the free-energy
difference $\Delta F_{\rm ab}=F_{\rm b}-F_{\rm a}$ is 
$\Delta F_{\rm ab} = (\gamma_{\rm a}-\gamma_{\rm b})\ln N =0.47\ln N$. Since the functional
form of $F_i$ and the  values of $\gamma_{\rm a}$ and $\gamma_{\rm b}$ are strictly correct only in
the limit of very large $N$, the use of short polymer lengths here ($N$=20--200) is expected
to lead to finite-size effects that may alter the form for $F_i$ and/or the effective scaling exponents.
In addition, to our knowledge the value of $\gamma_{\rm c}$ has not been measured. For these reasons
it is of interest to calculate $F_{\rm a}$, $F_{\rm b}$ and $F_{\rm c}$ directly for the range of $N$
relevant to this study.

\begin{figure}[!ht]
\begin{center}
\vspace*{0.2in}
\includegraphics[width=0.40\textwidth]{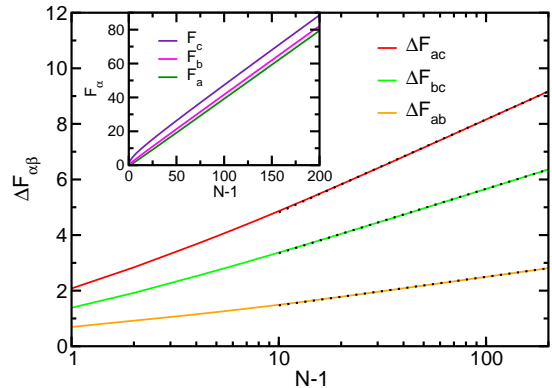}
\end{center}
\caption{Free-energy differences vs polymer length calculated from PERM simulations. 
These differences are defined $\Delta F_{\rm ab}\equiv F_{\rm b}-F_{\rm a}$,
$\Delta F_{\rm ac}\equiv F_{\rm c}-F_{\rm a}$, and $\Delta F_{\rm bc}\equiv F_{\rm c}-F_{\rm b}$,
where (i) $F_{\rm a}$ is the free energy of a free polymer; (ii) $F_{\rm b}$ is that for a
polymer tethered to a hard flat wall and (iii) $F_{\rm c}$ is that for a polymer tethered to a 
corner of an infinitely large confining cube. The dashed lines overlaid on the curves are fits
to the function $\Delta F_{\lambda\mu}=C_{\lambda\mu}+\alpha_{\lambda\mu}\ln N$, where
$C_{\rm ac}=1.476\pm 0.004$, $\alpha_{\rm ac}=1.451\pm 0.001$,
$C_{\rm bc}=1.031\pm 0.002$, $\alpha_{\rm bc}=1.006\pm 0.001$, and
$C_{\rm ab}=0.444\pm 0.002$, $\alpha_{\rm ab}=0.446\pm 0.002$.
The inset shows the free energies $F_{\rm a}$, $F_{\rm b}$ and $F_{\rm c}$ vs $N$.  }
\label{fig:delF.gamma}
\end{figure}

We employ the pruned-enriched-Rosenbluth method (PERM) to directly calculate $F_{\rm a}$, 
$F_{\rm b}$ and $F_{\rm c}$, which then yield all three free-energy differences. The 
PERM simulations are implemented in the same manner described in Ref.~\onlinecite{tree2013dna}.
Figure~\ref{fig:delF.gamma} shows the variation of $\Delta F_{\rm ab}$, $\Delta F_{\rm ac}$
and $\Delta F_{\rm bc}$ with polymer length. The inset shows $F_{\rm a}$, $F_{\rm b}$
and $F_{\rm c}$ used to calculate the differences. The dotted curves in the main figure
overlaid on the data are fits to the function $\Delta F_{\lambda\mu}=C_{\lambda\mu} 
+ \alpha_{\lambda\mu} \ln N$ in the range $N$=$10$--$200$. We find that the best-fit
parameter values of 
$C_{\rm ac}=1.476\pm 0.004$, $\alpha_{\rm ac}=1.451\pm 0.001$,
$C_{\rm bc}=1.031\pm 0.002$, $\alpha_{\rm bc}=1.006\pm 0.001$, and
$C_{\rm ab}=0.444\pm 0.002$, $\alpha_{\rm ab}=0.446\pm 0.002$. The values for 
$\Delta F_{\rm ab}$ compare well with the values obtained using the SCH method in 
Fig.~\ref{fig:F.scale}. 
%
%

Checking the validity of the expressions for $\alpha_{\rm ac}$ and $\alpha_{\rm bc}$ requires
a value for $\gamma_{\rm c}$, the exponent for a corner-tethered polymer. We are unaware of
any calculation of this exponent.  However, we do note that in Refs.~\onlinecite{maghrebi2011entropic} 
and \onlinecite{maghrebi2012polymer} the quantity 
$\alpha_{\rm bc^\prime} \equiv \gamma_{\rm c^\prime}-\gamma_{\rm b}$ 
was calculated using lattice Monte Carlo simulations for self-avoiding walk, where 
$\gamma_{\rm c^\prime}$ is the exponent for a polymer whose end is tethered to a tip of a 
hard conical object. Choosing a cone half-angle $\Theta=0.77\pi >\pi/2$ means that
the tethered polymer confined to lie in a conical subspace with a half-angle of 0.23$\pi$, 
which has a solid angle equal to that of the corner-tethered system of $\Omega=\frac{1}{8}(4\pi)$. 
From Fig.~3 of Ref.~\onlinecite{maghrebi2011entropic} the exponent difference is estimated to be
$\alpha_{\rm ac}\equiv \gamma_{\rm c}-\gamma_{\rm a} \approx 1.4$. This is close to the value 
of 1.451 obtained from the PERM simulations for a corner-tethered polymer.

\section{Free energy versus end-monomer position from a flat wall}
\label{app:b}

\begin{figure}[!ht]
\begin{center}
\vspace*{0.2in}
\includegraphics[width=0.40\textwidth]{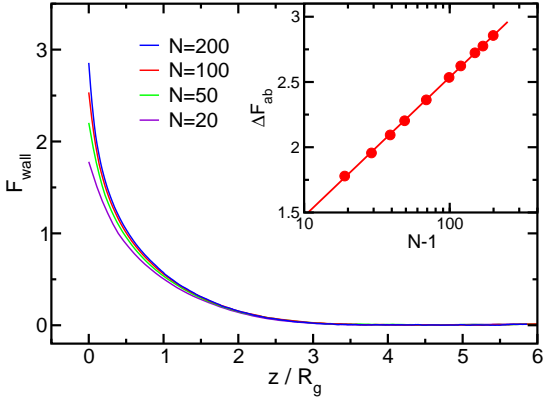}
\end{center}
\caption{Conformational free energy of a polymer near a flat wall $F_{\rm wall}$ vs scaled 
distance of an end monomer away from the wall, $z/R_{\rm g}$. Results are shown for different
polymer lengths. The inset shows the difference $\Delta F_{\rm ab}\equiv F_{\rm wall}(0)-
F_{\rm wall}(\infty)$ vs $N$. The solid line shows the fit to $\Delta F_{\rm ab}=
C_{\rm ab}+\alpha_{\rm ab}\ln N$, which yields $\alpha_{\rm ab}=0.463\pm 0.003$ and $C_{\rm ab} 
=0.40\pm 0.01$.  }
\label{fig:F.scale}
\end{figure}

In this appendix we calculate the variation of the conformational free energy with the distance $z$ 
of the end monomer from an infinitely wide hard flat wall. We employ the self-consistent histogram 
(SCH) method,\cite{frenkel2002understanding} which we have previously used to calculate free
energy functions for other processes such as polymer translocation.\cite{polson2013simulation}
Figure~\ref{fig:F.scale} shows the variation of the free energy of a 
polymer as a function of the distance of one end monomer away from the wall. Results are shown for 
polymers of length $N$=20, 50, 100 and 200, and the distance is scaled with respect to $R_{\rm g}$,
the radius of gyration of a free polymer.  As expected, the free energy rises monotonically as 
the polymer approaches the wall as a result of a decrease in conformational entropy. 
In addition, the free energy levels off to a constant value at distances far from the wall.
Upon approaching the wall, the distance at which the free energy becomes appreciable relative
to $k_{\rm B}T$ is $z\approx R_{\rm g}$.  We note that near the wall the free energy increases 
slightly with increasing polymer length.  Note that the quantity
$\Delta F_{\rm ab}\equiv F_{\rm wall}(0)-F_{\rm wall}(\infty)$, is the free-energy
difference between a free and a wall-tethered polymer. The inset of the figure shows
$\Delta F_{\rm ab}$ vs $N$ for the range $N=20$--$200$. The solid line is a fit to the 
function $\Delta F_{\rm ab} = C_{\rm ab} + \alpha_{\rm ab} \ln N$, which yielded best-fit parameter 
values of $\alpha_{\rm ab} = 0.463\pm 0.003$ and $C_{\rm ab}=0.40\pm 0.01$. The value for 
$\alpha_{\rm ab}$ compares well with the predicted value of 0.47 and the value of the 
$0.446\pm 0.002$ measured using PERM simulations in Appendix~\ref{app:a}. 
%
%

\section{Derivation of equations for the theoretical model}
\label{app:c}

In this appendix we develop the theoretical model used to understand the scaling
of the localization free energy with $N$ and confining cavity size, as well as the
cavity shape and location of the localization point on the cavity surface.
We first develop an approximation for the free-energy cost of localizing an end monomer
to a point on the wall of the confining cube located at the center of one of the cube faces. 
In practice, the validity of the derived result does not require that it be exactly
at the center, but rather only that it not be too close to one of the corners of the cube.

\begin{figure}[!ht]
\begin{center}
\vspace*{0.2in}
\includegraphics[width=0.40\textwidth]{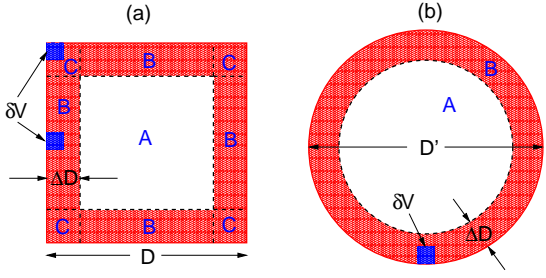}
\end{center}
\caption{Illustration of the various quantities used for the theoretical model.
(a) Region~A is cubic subspace of the confining cube. Region~C is composed of eight
small cubes near the corners of the confining cube, and region~B is the remaining
subspace located near the walls of the cube.  When a polymer is located in region~A,
it does not interact with the walls of the box. However, when it is located in regions B or C,
it does interact, leading to an increase in the conformational entropy, as explained
in the text. (b) As in panel (a), except for a confining sphere. There is only a single region
(B) where the polymer interacts with the confining wall. Note that $D^\prime$
=$(4\pi/3)^{1/3}D$, as explained in the text.}
\label{fig:box}
\end{figure}

Consider a cubic box of dimension $D$ and volume $V=D^3$ within which a polymer is confined. 
The probability distribution for the position of an end monomer ${\bf R}_0$ is
\begin{eqnarray}
P({\bf R}_0) = \frac{e^{-\beta F({\bf R}_0)}}{\int_V d^3{\bf R}_0 e^{-\beta F({\bf R}_0)}},
\label{eq:PR0}
\end{eqnarray}
where $F({\bf R}_0)$ is the conformational free energy of the polymer for this end-monomer 
position.  If the confining cube is sufficiently large relative to the average polymer size
then there exists an interior region of the cube where a polymer located inside does not appreciably 
interact with the walls of the box. Let us call this region A and its volume $V_{\rm a} (<V)$. If 
the end monomer lies within this subspace, we expect the conformational free energy of the polymer 
to be otherwise independent of position.  Call this free energy  $F_{\rm a}$. Outside this 
subspace is a layer of width $\Delta D\approx R_{\rm g}$ adjacent to the walls of the box. 
Let us write $\Delta D = \mu R_{\rm g}$, where the dimensionless
scaling factor $\mu$ is of order unity. We divide this region into two smaller regions. Region~C 
is comprised of the eight cubes, each of side length $\mu R_{\rm g}$, that lie in the corners
of the confining cube.  Region B is composed of the six regions near the wall that lie further away
from the corners of the cube.  If the end monomer lies at a position inside regions B and C
the polymer interacts with the walls of the box and the reduction of the conformational 
entropy causes the free energy to be greater than $F_{\rm a}$. The free energy
will of course depend on the exact location of the end monomer inside this layer, i.e.
how close it is to the wall or corner. For simplicity, we model this effect by choosing
a free energy that is taken to be constant in each subspace, which we label $F_{\rm b}$
and $F_{\rm c}$ for regions B and C, respectively. Figure~\ref{fig:box}(a) illustrates the 
regions of the confining cube.

Consider the case where the end monomer lies in region~B. From Eq.~(\ref{eq:PR0}) it follows
that the probability distribution for ${\bf R}_0$ is then
\begin{eqnarray}
P_{\rm b}({\bf R}_0) = \frac{e^{-\beta F_{\rm b}}}{V_{\rm a} e^{-\beta F_{\rm a}} 
     + V_{\rm b} e^{-\beta F_{\rm b}} 
               + V_{\rm c} e^{-\beta F_{\rm c}}} ~~~{\rm ({\bf R}_0~in~B)}.\nonumber\\
\end{eqnarray}
The probability that the end monomer lies in a smaller subspace of region B of volume
$\delta V$ is determined by integration: $P(\delta V) = \int_{\delta V} d^3{\bf R}_0 P({\bf R}_0)$. 
It can be shown that this leads to
\begin{eqnarray}
P_{\rm b}(\delta V) = \frac{(\delta V/V)}{(V_{\rm a}/V) e^{\beta \Delta F_{\rm ab}} + (V_{\rm b}/V)
                + (V_{\rm c}/V) e^{\beta \Delta F_{\rm cb}}}, \nonumber\\
\end{eqnarray}
where $\Delta F_{\rm ab} \equiv F_{\rm b}-F_{\rm a}$ and $\Delta F_{\rm cb} \equiv F_{\rm b}-F_{\rm c}$. 

The total free-energy cost of confinement of the end monomer to $\delta V$ is 
$F_{\rm conf} = -k_{\rm B}T\ln P_{\rm b}(\delta V)$. Consequently,
\begin{eqnarray}
F_{\rm conf} = k_{\rm B}T\ln(V/\delta V) + \Delta F_{\rm loc},
\end{eqnarray}
where the first term is equivalent to the free-energy cost of localizing a hypothetical detached 
end monomer. The second term, defined
\begin{eqnarray}
\Delta F_{\rm loc}/k_{\rm B}T \equiv \ln \left[ \left({V_{\rm a}}/{V}\right) e^{\beta \Delta F_{\rm ab}} 
                         + \left({V_{\rm b}}/{V}\right) + \left({V_{\rm a}}/{V}\right) 
                           e^{\beta \Delta F_{\rm cb}}\right], \nonumber \\
\label{eq:Fint_app}
\end{eqnarray}
accounts for the attachment of the end monomer to the rest of the polymer, whose conformational
entropy is determined by the end-monomer location. This, of course, is the approximation for
the chain-end localization free energy.
Note that in the limit of very large confinement volume the boundary layer regions B and C become
negligible by comparison, i.e. $V_{\rm a}/V\rightarrow 1$, $V_{\rm b}/V\rightarrow 0$, and 
$V_{\rm c}/V\rightarrow 0$, and thus $\Delta F_{\rm loc} = \Delta F_{\rm ab}$, as expected.

Next, we make the following approximations for $\Delta F_{\rm ab}$  and 
$\Delta F_{\rm cb}$ (=$-\Delta F_{\rm bc}$).  First note that the quantities $F_{\rm a}$, 
$F_{\rm b}$ and $F_{\rm c}$ are the conformational free energies of the polymer (assumed 
constant) for end-monomer position in each or the three regions. Approximating each of these as 
the free energies of a free polymer ($F_{\rm a}$) or an end-tethered  polymer to a large 
flat wall or corner ($F_{\rm a}$ and $F_{\rm c}$, respectively) we write 
\begin{eqnarray}
\Delta F_{\rm \lambda\mu}/k_{\rm B}T & = & C_{\lambda\mu} + \alpha_{\lambda\mu}\ln N
\label{eq:DFmn}
\end{eqnarray}
and use the values of $C_{\rm ab}$, $C_{\rm cb}$, $\alpha_{\rm ab}$ and $\alpha_{\rm cb}$
obtained from the PERM simulations in Appendix~\ref{app:a}.  Note that the use of the scaling
results from Appendix~\ref{app:a} is strictly correct only for very large confinement cubes 
($R_{\rm g} \ll D$) and for end-monomer locations that lie on the confining surface, rather than 
at arbitrary position in regions B and C, as assumed here.

Finally, let us define the ratio $r\equiv D/R_{\rm g}$, where $R_{\rm g}$ is the radius of gyration.
Since the subspaces B and C constitute a layer of width $\Delta D = \mu R_{\rm g}$ near the walls
of the cube, the volume $V_{\rm a}$ is given by $V_{\rm a}=(D-2\mu R_{\rm g})^3$. In addition,
$V_{\rm c} = 6(\mu R_{\rm g})^3$ and $V_{\rm b} = V-V_{\rm a}-V_{\rm b}$. It follows that
\begin{eqnarray}
V_{\rm a}/V & = & 1 - 6\mu r^{-1} + 12\mu^2 r^{-2} -8\mu^3 r^{-3}  \nonumber \\
V_{\rm b}/V & = & 6\mu r^{-1} - 12\mu^2 r^{-2}  \nonumber \\
V_{\rm c}/V & = & 8\mu^3 r^{-3}.
\label{eq:Vabc}
\end{eqnarray}

The conformational free-energy cost of localizing an end monomer to a point near a wall, 
$\Delta F_{\rm loc}$ is approximated using Eqs.~(\ref{eq:Fint_app})--(\ref{eq:Vabc}). We use 
these equations to estimate the free energy of localization to a point in the middle of one 
of the cube faces.  For the calculations used to generate the results in 
Figs.~\ref{fig:theory.cube.sphere} and \ref{fig:theory.N} we use a value for the dimensionless 
factor $\mu$ appearing in Eqs.~(\ref{eq:Vabc}) of $\mu$=0.7. The results of Fig.~\ref{fig:F.scale} 
in Appendix~\ref{app:b} demonstrate that this is a reasonable choice.  Using other values of order 
unity have a small quantitative effect on the predictions, but the qualitative trends are unaltered.

Next we consider end-monomer localization near a corner of the box. If the end monomer lies in 
region~C the probability distribution for ${\bf R}_0$ is
\begin{eqnarray}
P({\bf R}_0) = \frac{e^{-\beta F_{\rm c}}}{V_{\rm a} e^{-\beta F_{\rm a}} + V_{\rm b} e^{-\beta F_{\rm b}}
             + V_{\rm c} e^{-\beta F_{\rm c}}}.
\end{eqnarray}
The probability that ${\bf R}_0$ lies in a small volume $\delta V$ located in region $C$ 
is obtained from integration, $P_{\rm c}(\delta V) = \int_{\delta V} P({\bf R}_0) d^3{\bf R}_0$. 
It can then be shown that the total free-energy cost of end-monomer confinement to $\delta V$ is 
$F_{\rm conf}=-k_{\rm B}T\ln P_{\rm c}(\delta V) = k_{\rm B}T\ln(V/\delta V) + \Delta F_{\rm loc}$, 
where the end-monomer localization free energy is given by
\begin{eqnarray}
\Delta F_{\rm loc}/k_{\rm B}T = \ln\left[ (V_{\rm a}/V)e^{\beta \Delta F_{\rm ac}} +
                    (V_{\rm b}/V)e^{\beta \Delta F_{\rm bc}} + (V_{\rm c}/V)  \right],
\nonumber \\
\label{eq:DFintcor}
\end{eqnarray}
where $\Delta F_{\rm ac}\equiv F_{\rm c}-F_{\rm a}$ and $\Delta F_{\rm bc}\equiv F_{\rm c}-F_{\rm b}$.
These free-energy differences are approximated once again using Eq.~(\ref{eq:DFmn}),
where the quantities $C_{\lambda\mu}$ and $\alpha_{\lambda\mu}$ are obtained from the fits
to the PERM simulation data of Appendix~\ref{app:a}.

The free-energy cost for end-monomer localization to a corner of the cube is approximated using 
Eqs.~(\ref{eq:DFmn}), (\ref{eq:Vabc}) and (\ref{eq:DFintcor}). As noted above, to generate the 
results in Figs.~\ref{fig:theory.cube.sphere} and \ref{fig:theory.N} we use a value for the scaling 
factor of $\mu$=0.7. 

Finally, let us consider the case of end-monomer localization a point on the wall of a confining
sphere. Unlike the case of the cube, we need only define a single region for which the polymer
interacts with the confining wall. We denote the interior region A and the outer region B,
as illustrated in Fig.~\ref{fig:box}(b). Following the same approach as above, it is easy to
show that the localization free energy is given by
\begin{eqnarray}
\Delta F_{\rm loc}/k_{\rm B}T \equiv \ln \left[ \left({V_{\rm a}}/{V}\right) 
         e^{\beta \Delta F_{\rm ab}} + \left({V_{\rm b}}/{V}\right)\right], ~~~~
\label{eq:DFintsph}
\end{eqnarray} 
where 
\begin{eqnarray}
V_{\rm a}/V & = & 1 - 6\mu (c/r) + 12\mu^2 (c/r)^2 -8\mu^3 (c/r)^3  \nonumber \\
V_{\rm b}/V & = & 6\mu (c/r) - 12\mu^2 (c/r)^2 +8\mu^3 (c/r)^3, ~~~~~~
\label{eq:Vab}
\end{eqnarray}
and where $c\equiv (\pi/6)^{1/3}$. Note that since $D\equiv V^{1/3}$, it follows that the 
confining sphere diameter $D^\prime$ is given by $D^\prime=(\pi/6)^{1/3}D$, which is
the origin of the factor $c$ in these equations. As before $F_{\rm ab}$ is determined by
Eq.~(\ref{eq:DFmn}), and we use $\mu$=0.7.

%


\end{document}